\documentclass[usenatbib,useAMS]{mn2e}
\usepackage{epsfig}
\usepackage{times}
\usepackage{amsmath}

\usepackage{verbatim}

\newcommand{\simgt}%
        {\,\hbox{\lower0.6ex\hbox{$\sim$}\llap{\raise0.6ex\hbox{$>$}}}\,}
\newcommand{\simlt}%
        {\,\hbox{\lower0.6ex\hbox{$\sim$}\llap{\raise0.6ex\hbox{$<$}}}\,}
\usepackage{color}
\title[]{2D simulations of the double-detonation model for thermonuclear transients from low-mass carbon--oxygen white dwarfs}
\author[]{S. A. Sim$^1$, M. Fink$^{2,3}$, M. Kromer$^{3}$, F. K. R\"{o}pke$^{2,3}$, A. J. Ruiter$^{3}$, W. Hillebrandt$^{3}$\\
$^{1}$Research School of Astronomy and Astrophysics, Mount Stromlo Observatory,
Cotter Road, Weston Creek, ACT 2611, Australia\\ 
$^{2}$Universit{\"a}t W{\"u}rzburg, Emil-Fischer-Str.~31, 97074 W{\"u}rzburg, Germany\\
$^{3}$Max-Planck-Institut f\"{u}r Astrophysik,
Karl-Schwarzschildstr.~1, 85748 Garching, Germany
}

\date{9 November 2011}

\pagerange{\pageref{firstpage}--\pageref{lastpage}}
\pubyear{}
\begin{document}
\maketitle
\label{firstpage}

\begin{abstract}
Thermonuclear explosions may arise in binary star systems in which a carbon--oxygen (CO) white dwarf (WD) accretes helium-rich material from a companion star. If the accretion rate allows a sufficiently large mass of helium to accumulate prior to ignition of nuclear burning, the helium surface layer may detonate, giving rise to an astrophysical transient. Detonation of the accreted helium layer generates shock waves that propagate into the underlying CO WD\@. 
This might directly ignite a detonation of the CO WD at its surface (an edge-lit secondary detonation) or compress the core of the WD sufficiently to trigger a CO detonation near the centre. If either of these ignition mechanisms works, 
the two detonations (helium and CO) can then release sufficient energy to completely unbind the WD\@.
These ``double-detonation'' scenarios for thermonuclear explosion of WDs have previously been investigated as a potential channel for the production of type~Ia supernovae from WDs of around one solar mass. Here we extend our 2D studies of the double-detonation model to significantly less massive CO WDs, the explosion of which could produce fainter, more rapidly evolving transients. We investigate the feasibility of triggering a secondary core detonation by shock convergence in low-mass CO WDs and the observable consequences of such a detonation. Our results suggest that core detonation is probable, even for the lowest CO core masses that are likely to be realized in nature. To quantify the observable signatures of core detonation, we compute spectra and light curves for models in which either an edge-lit or compression-triggered CO detonation is assumed to occur. We compare these to synthetic observables for models in which no CO detonation was allowed to occur.
If significant shock compression of the CO WD occurs prior to detonation, explosion of the CO WD can produce a sufficiently large mass of radioactive iron-group nuclei to significantly affect the light curves. {In particular, this can lead to relatively slow post-maximum decline.} If the secondary detonation is edge-lit, however, the CO WD explosion primarily yields intermediate-mass elements that affect the observables more subtly. In this case, near-infrared observations and detailed spectroscopic analysis would be needed to determine whether a core detonation occurred. We comment on the implications of our results for understanding peculiar astrophysical transients including SN~2002bj, SN~2010X and SN~2005E.
\end{abstract}


\section{Introduction}
\label{sect_intro}

Type~Ia supernovae (SNe~Ia) are understood to result from the
thermonuclear disruption of a carbon--oxygen (CO) white dwarf (WD) star
\citep[e.g.][]{hillebrandt00}. One possible mechanism for
igniting such an explosion can occur in binary systems in which a
primary CO WD accretes He-rich material from a donor star. When a
sufficiently large surface He layer is accreted, it is expected to
ignite explosively leading to detonation of the accreted He layer
{\citep[see e.g.][]{nomoto80,nomoto82,woosley80,woosley94}}. 
Detonation of the He
layer can then lead to a secondary detonation of the core, either by
directly igniting the CO fuel near to the interface with the overlying
He layer \citep[see e.g.][]{nomoto82, livne90b} 
or due to compressional heating of the core by inward
propagating shocks \citep[see e.g.][]{livne90a}. The consequence of this ``double-detonation'' model is the incineration of the CO WD and its He-rich outer layer, leading to an explosion in which the primary star is completely destroyed. 

However, the question of whether the secondary detonation forms is challenging owing to
the wide range of relevant length-scales that must be resolved if it is to be simulated \citep{seitenzahl09b,seitenzahl09a}. 
To date only a few 
multi-dimensional studies have been made of the double-detonation
scenario for a handful of progenitor models \citep{livne90b,
  dgani90a, livne91a, livne95a, benz97a, garcia99a,
  forcada06a, forcada07a, fink07,fink10}. If the secondary core detonation does not occur, the result of He ignition is quite different from the double-detonation SN~Ia model -- 
as described by \citet{bildsten07} and \citet{shen09}, explosive burning of an accreted He-layer alone will lead to a thermonuclear transient that is roughly ten times fainter and evolves significantly faster than a {SN~Ia}. 
Dubbed ``point-Ia'' (hereafter p-Ia), this class of explosion is readily accessible to observation by the current generation of transient surveys (e.g.\ the Panoramic Survey Telescope \& Rapid Response System [Pan-STARRS]\footnote{http://pan-starrs.ifa.hawaii.edu}, the Palomar Transient Factory [PTF]\footnote{http://www.astro.caltech.edu/ptf} and
planned wide-field surveys by instruments such as SkyMapper [\citealt{keller07}] and the Large Synoptic Survey Telescope [\citealt{LSST09}]). Indeed, transient events with some similarities to the predicted properties of p-Ia explosions have already been reported and modelled in the context of the p-Ia scenario {\citep[e.g.][]{foley09,perets10,poznanski10,kasliwal10,waldman11,sullivan11}}. It is to be expected that other similar events will be found and studied in the near future.

In our previous studies \citep{fink10,kromer10}, we investigated the
double-detonation scenario for systems with fairly massive CO cores
($M_{\mbox{\scriptsize CO}} > 0.81$~M$_{\odot}$). Such cases are the
most promising for yielding thermonuclear explosions as bright as normal SNe~Ia. 
In those works, we focused on the possibility of core detonation triggered by converging shocks deep in the CO WD\@. 
We found that secondary core detonation was very
likely in all of the models we considered \citep{fink10}. This implies that the p-Ia scenario should not be realised
following He detonation in such systems unless some additional effect
comes into play (see Section~\ref{sect:core_det}).
We now wish to extend our 2D studies of the double-detonation scenario
to investigate systems with less massive CO cores
(e.g.\ $M_{\mbox{\scriptsize CO}} \simlt 0.6$~M$_{\odot}$) and
quantify the observable properties of double detonations for such systems. 
Compared to their
more massive counterparts, these low-mass systems have two important
differences. First, prior to any explosion, the central density of the
CO core will be lower. In principle, this might make it harder for the
converging shocks to compress the centre sufficiently for a core
detonation to occur. Previous 2D studies have already suggested
that secondary detonations can be produced for CO cores with masses as
low as 0.55~M$_{\odot}$ \citep{livne95a}. 
Thus the first objective of
our study is to extend our studies to even lower mass, around the
minimum CO core mass that is expected to be
realized in nature 
(in the recent binary synthesis calculations of
\citealt{ruiter11}, the lowest CO core mass is ${\sim}0.45$~M$_{\odot}$). 
Second, even if a core detonation does occur, the low core density
means it will produce
little $^{56}$Ni \citep[see e.g.\ the lowest mass
model of ][which yields only 0.14~M$_\odot$ of $^{56}$Ni]{livne95a}.
In the limit of a very low $^{56}$Ni-yield in the core, the
radioactive products produced in the He-detonation may still play a dominant
role in determining the explosion brightness and light curve
evolution (as in the p-Ia scenario). Our second goal,
therefore, will be to quantify the observable properties of
explosions in which the core detonates but produces only a small mass ($\simlt 0.15$~M$_{\odot}$) of $^{56}$Ni.

We begin, in Section~\ref{sect:hydro}, by introducing the models adopted
for this study and the suite of numerical simulations used to study
them. In Section~\ref{sect:csdd} we describe our simulations in which we
investigate shock convergence and detonation in low mass CO cores. We
then present the results of alternative explosion simulations (in
which either an edge-lit CO core detonation is assumed or in which it is
assumed that no core detonation occurs; see Section~\ref{sect:othermods}).
In Section~\ref{sect:obs} we present the synthetic observables computed for all our simulations before discussing our results and drawing conclusions in Section~\ref{sect:discuss}.

\section{Methods}
\label{sect:hydro}

For this study we have performed sets of 2D numerical simulations that follow the explosion dynamics, nucleosynthesis and radiation transport of different explosions for two specific initial systems. Here we describe the parameters of the initial systems and the means by which the numerical simulations were performed.

\subsection{Initial models}
\label{sect:initial_models}

As initial conditions for our explosion simulations, we adopt models describing 
the state of the system immediately prior to He detonation.
These are not based on evolutionary/accretion calculations {\citep[cf.][]{woosley94,woosley11}} but are
idealized representations of low-mass CO WDs that have accreted a surface
layer of He. Thus, in this work, it is an assumption
that the system has evolved to reach conditions for He detonation --
we only determine whether subsequent detonation of
the CO core is probable and quantify its observable consequences.

We have considered two sets of system parameters, which are described below. {
Population synthesis calculations suggest that 
He-rich accretion by CO
WDs leading to helium shell detonation most commonly occurs in binaries
containing a CO WD primary and a He WD donor \citep[e.g.][]{ruiter11}\footnote{Note, however, that the Ruiter et al. (2011) calculations assumed explosions occurred once a 0.1 M$_{\odot}$ layer of He builds up; the systems we discuss here require additional He accretion beyond that point.}.
However, that channel is not
predicted to yield significant numbers of systems with very low CO mass
($\simlt 0.6~$M$_{\odot}$).  For the CO masses we consider a
non-degenerate He star companion is more probable \citep[see figure 3 of
][]{ruiter11}. The evolutionary path to a low-mass CO WD accreting from
a He star involves one or more stable mass-transfer episodes and at least one 
common envelope phase leading to a tight binary ($< 1$~hr orbital
period). Compared to binaries with He WD donors, systems that evolve to
a final state involving a CO WD and a He star donor are rarer
and the evolutionary timescale is
usually relatively short ($< 1$~Gy; \citealt{ruiter11}). Nevertheless,
the He-star donor scenario is likely to be the most
promising route to the low CO mass explosions that we will discuss.
}

For our standard system (hereafter ``S'' model), we 
adopted a CO core mass of $M_{\mbox{\scriptsize CO}} = 0.58$~M$_{\odot}$ and a mass for the accreted He layer of $M_{\mbox{\scriptsize He}} = 0.21$~M$_{\odot}$, yielding a total mass of $M_{\mbox{\scriptsize tot}} = 0.79$~M$_{\odot}$.
Here, $M_{\mbox{\scriptsize He}}$ is close to the 
minimum mass for detonation suggested by
\citet{bildsten07} and \citet{shen09} for our chosen value of $M_{\mbox{\scriptsize CO}}$.
This model naturally extends the study of \citet{fink10} into the
regime of physically plausible low-mass systems that might be realised
in systems where a primary CO WD accretes from a He-burning star (see Section~\ref{sect:future_work}).
Its masses are very similar to the least massive models considered in the studies of \citet{woosley94},  
\citet{livne95a} and \citet{shen10}, and also to model CO.55HE.2 of 
\citet{waldman11}. 

As a second case, we also considered an extremely low-mass
system. This model
(hereafter model ``L'') is designed to
robustly bracket the
low-mass end of the distribution of potential initial
systems.
The adopted CO core mass ($M_{\mbox{\scriptsize CO}} =
0.45$~M$_{\odot}$) lies at the lower
boundary of the distribution in the population
synthesis calculations of \citet{ruiter11}. 
We have also adopted a very low mass ($M_{\mbox{\scriptsize He}} = 0.21$~M$_{\odot}$) for the He layer when the explosion occurs. 
This is close to the most optimistic (i.e.\ lowest) estimate for
the mass of He needed for detonation, following the arguments of
\citet{shen09}\footnote{From figure~5 of \citet{shen09}, a
  minimum He-layer mass of ${\sim}0.2$~M$_\odot$ (for $M_{\mbox{\scriptsize CO}} =
0.45$~M$_{\odot}$) is required, if one adopts the condition that the dynamical timescale is one tenth of the local heating timescale when dynamical burning sets in.}.
Moreover, the evolutionary models of \citet{woosley11} imply that the conditions suggested by \citet{shen09} lead to He layer masses that are generally too small for detonation.
Thus, our L-model likely lies outside the regime in which He detonation is probable. 
However, we include it as an important
numerical experiment to test the limit of the
double-detonation model -- \emph{if} He detonation in this system leads to
CO core detonation, then it can be concluded that our method 
would predict secondary detonation 
for any combination of CO-core/He-layer mass for which He detonation
is realistic. Our L-model is similar to the lowest mass
model considered by \citet{waldman11} (CO.45HE.2).

The initial models are set up in exactly the same manner as the models described by \citet{fink10} by choosing appropriate values for the temperature and central density of the CO core ($T_\text{c}, \rho_\text{c}$), 
and the temperature and density at the base of the He layer ($T_\text{b},
\rho_\text{b}$). These parameters are listed in
Table~\ref{tab:init_model}.

In our S-model we aimed at achieving a bright explosion and therefore
assumed a cold shell (i.e.\ as dense as possible for a given mass).  
For degenerate
matter, the exact value of $T$ does not matter.  Thus, we simply
assumed a constant temperature $T = 5 \times 10^5$~K in the whole
WD\@.
Following \citet{shen09} and \citet{waldman11}, we adopted higher temperatures
for our L-model ($T = 1 \times 10^7$~K in the core and $T = 2 \times
10^8$~K at the base of the shell, decreasing adiabatically outwards).  
Higher temperatures reduce the density, leading to less complete burning and 
making it harder to trigger a
secondary core detonation. Thus, this choice maintains the status of our L-model as a fairly extreme test for the plausibility of secondary detonation by shock compression. 

For simplicity, the core is assumed to consist of
uniformly mixed $^{12}$C and $^{16}$O (equal parts by mass) and the surface layer is
assumed to be pure He. Following \citet{fink10}, the density profile
within the model is constructed by solving for hydrostatic equilibrium
conditions with the adopted value of  $T_\text{c}$, $\rho_\text{c}$, $T_\text{b}$ and
$\rho_\text{b}$ using the same equation of state adopted by \citet{fink07,fink10}.

\begin{table}
\caption{Parameters defining the initial model.}
\label{tab:init_model}
\begin{tabular}{ccc}
  \hline
  Parameter & Model S & Model L \\ \hline
    $T_\text{c}$ (K) &
    $5 \times 10^5$ &
    $1 \times 10^7$ \\
    $\rho_\text{c}$ ($10^6$~g~cm$^{-3}$) &
    8.5 &
    3.81 \\
    $T_\text{b}$ (K) &
    $5 \times 10^5$ &
    $2 \times 10^8$ \\
    $\rho_\text{b}$ ($10^6$~g~cm$^{-3}$) &
    1.3 &
    0.592 \\
    $M_\text{CO}$$^a$ ($M_\odot$) &
    0.58 &
    0.45 \\
    $M_\text{He}$$^a$ ($M_\odot$) &
    0.21 &
    0.21 \\
    $M_\text{tot}$$^a$ ($M_\odot$) &
    0.79 &
    0.66 \\
    \hline
\end{tabular}\\
$^a$ Note that the masses are not independent parameters but follow from $T_\text{c}$, $\rho_\text{c}$, $T_\text{b}$ and $\rho_\text{b}$ and the assumption of hydrostatic equilibrium (see Fink et al. 2010).
\end{table}

\subsection{Explosion simulations, nucleosynthesis and radiation transport}

In most respects, our simulations were performed in the same way as those described by \citet{fink10}. Therefore, we will only summarise the main points and highlight the small number of modifications to the numerical implementation adopted.

All the simulations presented here were carried out in 2D with
rotational symmetry about the $z$-axis. As in \citet{fink10}, we begin by
igniting a detonation at a single point in the He layer. As noted in 
Section~\ref{sect:initial_models}, that
such a He detonation ignites is a fundamental assumption of our
simulations. In all cases, we choose to ignite the He detonation 
at the base of the He layer on the positive $z$-axis. {In the absence of evolutionary calculations prior to explosion, this is the simplest choice for He ignition but we note that it is disfavourable for edge-lit secondary detonations (see Section~\ref{sect:ELDD}).}

\begin{figure}
\epsfig{file=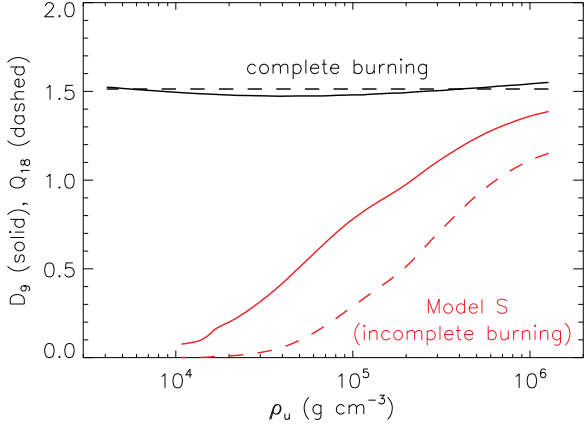,width=8cm}
\caption{Detonation speed ($D_{9}$; units of $10^9$~cm~s$^{-1}$)
  relative to unburnt matter (solid lines) and $Q$-values (energy
  release per unit mass in $10^{18}$~ergs~g$^{-1}$; dashed lines) for our
  S-model. The red curves indicate the final values obtained by our
  iterative calibration of the lateral detonation. The black curves
  show the values expected for a Chapman-Jouguet detonation if it were
  assumed that complete burning to $^{56}$Ni occurs at all densities.}
\label{fig:vdet}
\end{figure}

As in \citet{fink10}, detonations in CO and He
were modelled with a front tracking scheme using tabulated values for
both detonation speeds ($D$) and energy release per unit mass ($Q$) behind the burning front
(hereafter referred to as ``detonation tables'').  Since it takes place
entirely in a low-density incomplete burning regime, the He
detonation nucleosynthesis is very sensitive to the input parameters of
the front-tracking scheme.  The detonation tables were therefore
determined for each model of this work by applying the
hydrodynamics/post-processing iteration scheme described in the
appendix of \citet{fink10}.  This time, however, the setup in the
calibration runs was identical to that of the models, i.e., a
detonation propagating laterally through the same He layers.  
To calibrate the He-detonation speed, the Rankine-Hugoniot jump conditions for
detonations were solved for the minimum possible value of $D$, which
corresponds to the flow velocity $u_\text{ash}$ of the final ash state being exactly
sonic relative to the front.  This procedure was repeated for every point
on the tabulated grid of density values.  In the calculations we use
the same equation of state as in the
hydrodynamics code and the energy release from the previous iteration
step. This procedure leads to converged detonation speeds after
around six iteration steps.
The final detonation tables for our standard (``S'') model 
are illustrated in Figure~\ref{fig:vdet}. It can be seen that
complete He burning is never achieved for the densities present in the
model and that this substantially reduces the detonation speed at low
densities, compared to the complete burning case. 

Unlike in the models of \citet{fink10}, we suppressed any 
volume burning in the He layer before the arrival
of the detonation wave.
This provides
a well-defined initial condition in the He layer (allowing the
composition to change due to volume burning prior to detonation would not be self-consistent as we
do not simulate the evolution of the progenitor before the initiation
of the He detonation). 

We performed detailed nucleosynthesis calculations for the explosion models
using a tracer particle method \citep{travaglio04}. Here, tracer
particles are passively advected in the hydrodynamical simulation and
used to record the thermodynamic trajectories of mass elements. These are then used as input to a post-processing step in which detailed isotopic yields are obtained from calculations with
an extensive nucleosynthesis network (384 species). 
We adopted an updated version of the REACLIB reaction
rate library \citep[][updated 2009]{rauscher2000a} and a refined
method with variable tracer masses
\citep{seitenzahl10a} was applied.  Variable tracer masses allow for
a better spatial resolution at the edge of the CO core.

The nucleosynthesis tracer particles are used to reconstruct the detailed abundances throughout the ejecta.
This gives a complete, 2D model for the structure of the ejecta at the final time of the hydrodynamical simulations (i.e.\ density and composition as functions of expansion velocity in the ${r}$- and ${z}$-directions). We then used the {\sc artis} code \citep{sim07,kromer09} to compute synthetic light curves and spectra for the models. For all the radiative transfer simulations we used our set of atomic data extracted from CD23 of \citet{kurucz95} (see \citealt{kromer09}), but we expanded the range of ions included to {\sc i} -- {\sc vii} for elements with atomic number $22 < Z < 28$ to allow for higher ionization states that may be present at early times when the ejecta are hot.

\section{Secondary detonation by converging shocks; CSDD models}
\label{sect:csdd}

Our first simulations are the most similar to those described by
\citet{fink10} -- they study the shock convergence and potential
for formation of a secondary detonation via compression of the CO core 
in less-massive systems. 
We investigate this scenario for both our initial systems (S and L, see above).

As in \citet{fink10}, we simulate the propagation of the He detonation
as it wraps around the CO core. The He detonation drives a shock front that propagates into the
core leading to strong compression around a convergence
point. Although the convergence point is off-centre in both models, it
is \emph{less} off-centre than in the models of \citet{fink10}. This is a
continuation of the trend for ignition closer to centre in less
massive cores (see table~2 of \citealt{fink10}). 

Our first question is whether this compression
leads to a large enough volume of sufficiently hot and dense material that a
secondary CO detonation could ignite. To assess this we
compared the density and temperature reached in our simulation to
critical temperatures and densities for detonation from
\citet{niemeyer97} and \citet{roepke07}\footnote{Note that for the high densities reached in the shock
  convergence region, the critical volume for initiation of a
  detonation is small compared to our grid resolution \citep[see table
  1 of ][]{fink07}. 
  Therefore only
  temperature and density can be considered as detonation criteria in this case.}.
 From this comparison (see
Table~\ref{tab:hotspot}), we find that critical conditions for CO core
detonation are robustly met in the simulation for both our initial
systems -- in fact, the peak temperatures and densities are very
similar to those of models 1 and 2 of
\citet{fink10}, implying that core detonation is not significantly
harder to achieve in the systems we consider here. Given the extreme
properties of our model~L, we therefore conclude that 2D converging
shock simulations {performed with our current approach}
will favour secondary
core detonations for \emph{any} physically plausible pair of CO core/He-layer 
mass. {To study this further would require much higher resolution simulations that resolve the critical volumes for detonation.}

\begin{table}
  \caption{Conditions at the hot spot in our converging shock double-detonation models.
    $t_\text{ign}$ is the time at which critical conditions for
    ignition of the CO core detonation are reached while
    $z_\text{ign}/R_\text{CO}$ is the position of the hot spot (which lies on
    the $z$-axis of the simulation) in units of the CO core radius.  $T_\text{ign}$ and
    $\rho_\text{ign}$ are the temperature and density at the hot
    spot at $t = t_\text{ign}$.  $\Delta$ is the grid resolution.}
  \label{tab:hotspot}
  \begin{tabular}{ccc}
    \hline
    Parameter & Model S & Model L \\
    \hline
    $t_\text{ign}$ (s) &
    1.34 & 1.81 \\
    $z_\text{ign}$ ($10^8$~cm)&
    $-1.39$ & $-1.61$ \\
    $z_\text{ign}/R_\text{CO}$ &
    0.31 & 0.31 \\
    $T_\text{ign}$ ($10^9$~K) &
    6.44 & 4.68 \\
    $\rho_\text{ign}$ ($10^7$~g~cm$^{-3}$)&
    18.0 & 7.83 \\
    $\Delta$ ($10^6$~cm) &
    3.71 & 4.94 \\
    \hline
  \end{tabular}
\end{table}

Since our simulations of the shock convergence suggests that a
detonation in the CO core is likely, we initiate a
second detonation wave at the location of the hot spot in the CO core.
This detonation sweeps over the whole CO core and releases sufficient
energy to completely unbind the star. Hereafter, we will refer to the
results of these simulations as our converging-shock double-detonation (CSDD)
models (CSDD-S and CSDD-L for our two initial systems, respectively).

\begin{figure*}
\epsfig{file=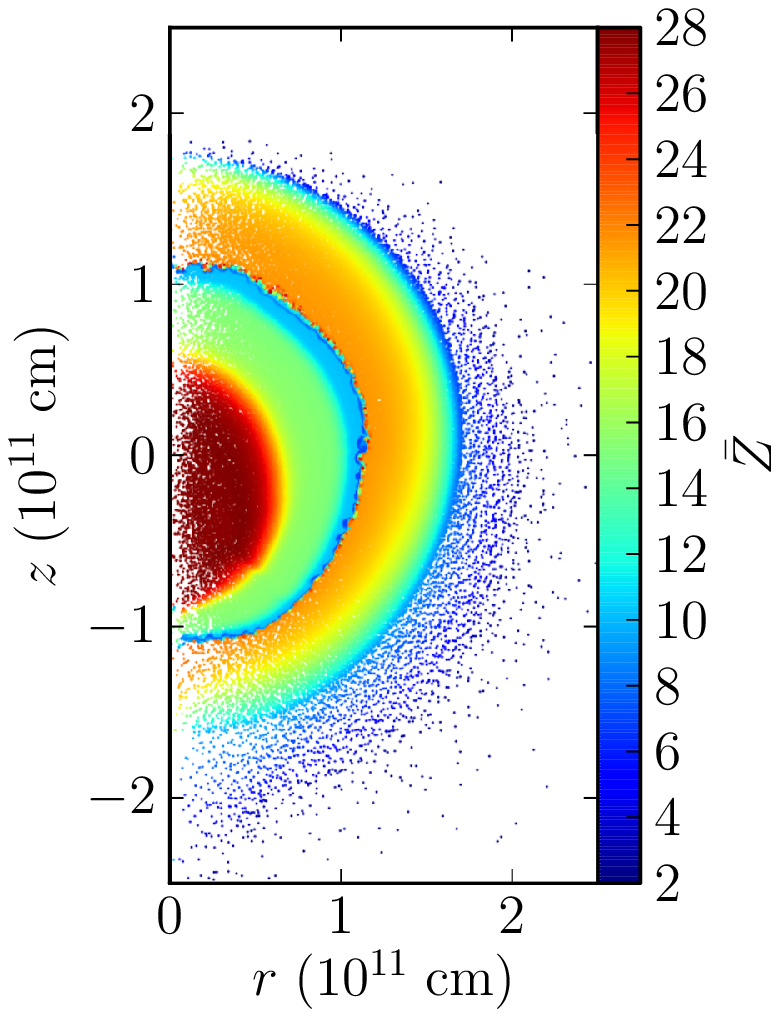, width=4cm}
\hspace{0.7cm}
\epsfig{file=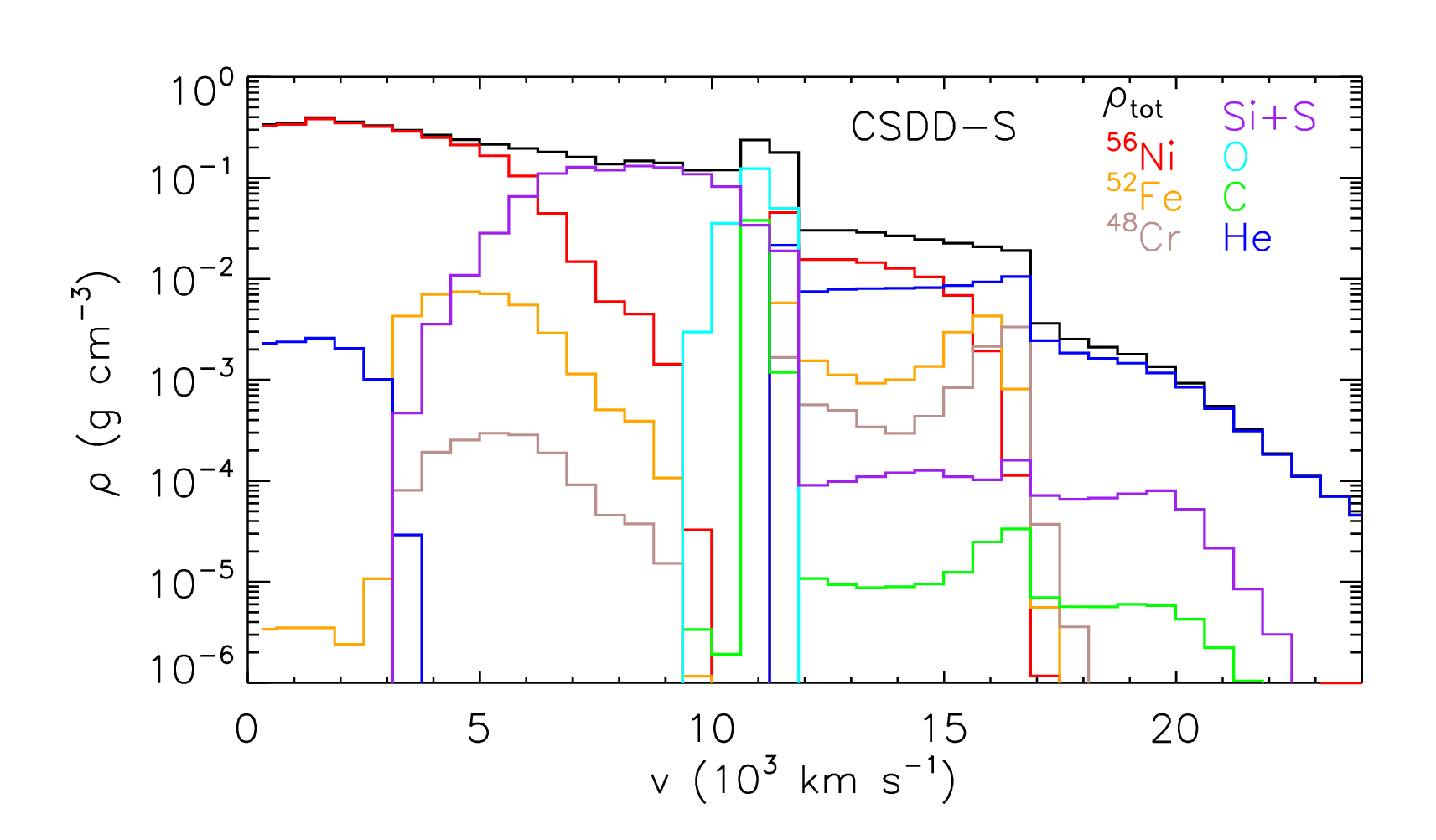,angle=90,width=9cm}\\
\vspace{0.3cm}
\epsfig{file=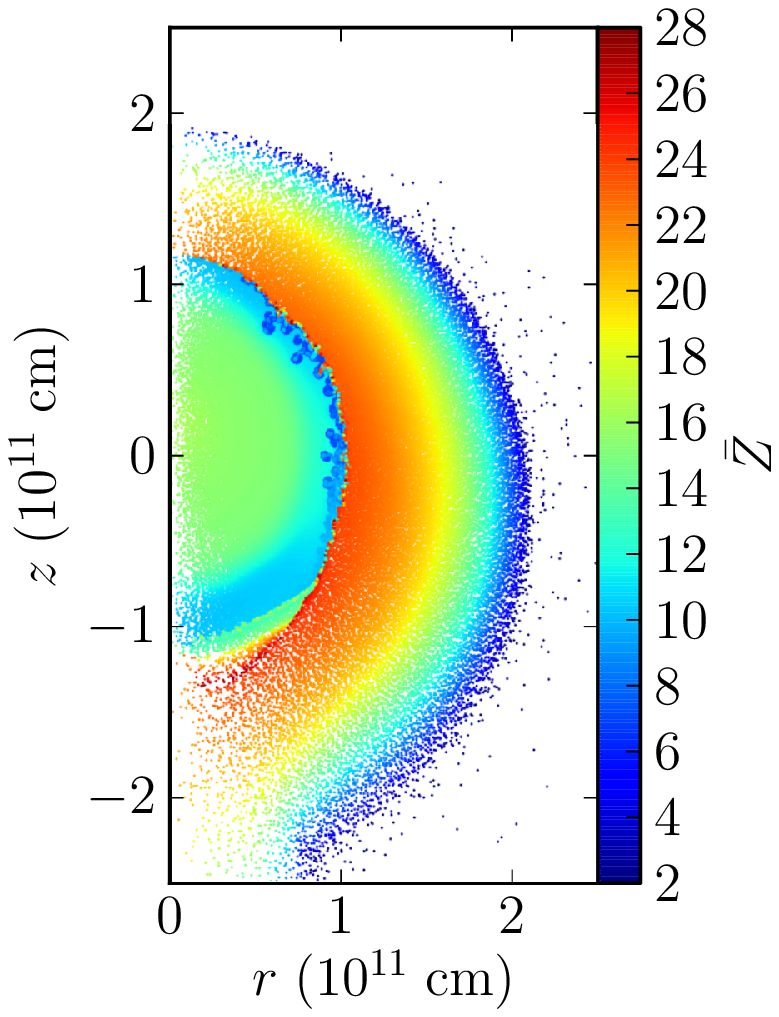, width=4cm}
\hspace{0.7cm}
\epsfig{file=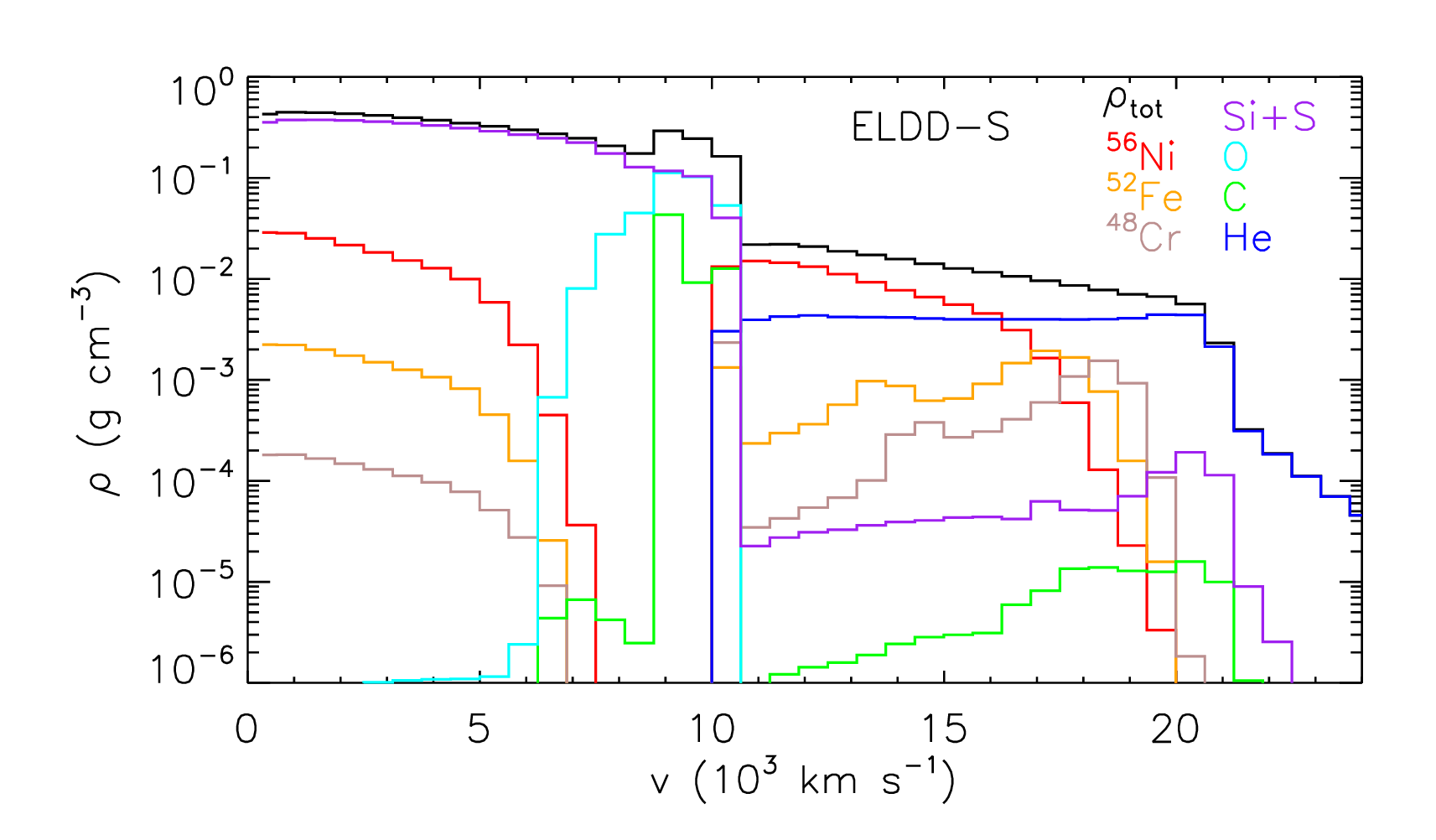,angle=90,width=9cm}\\
\vspace{0.3cm}
\epsfig{file=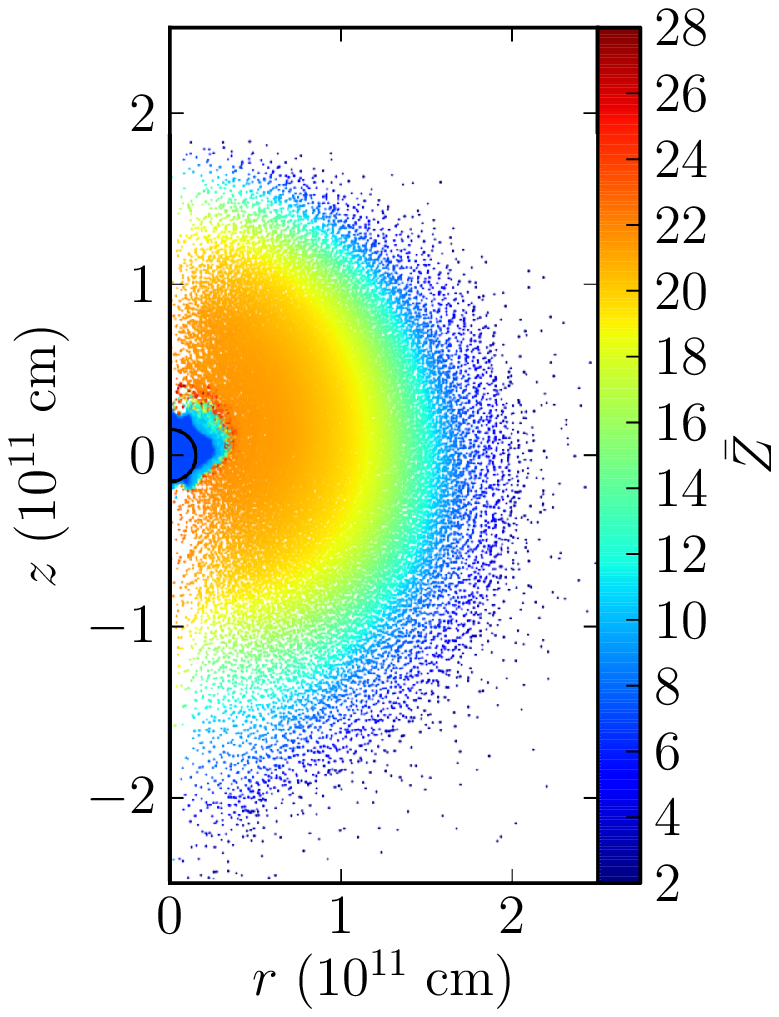, width=4cm}
\hspace{0.7cm}
\epsfig{file=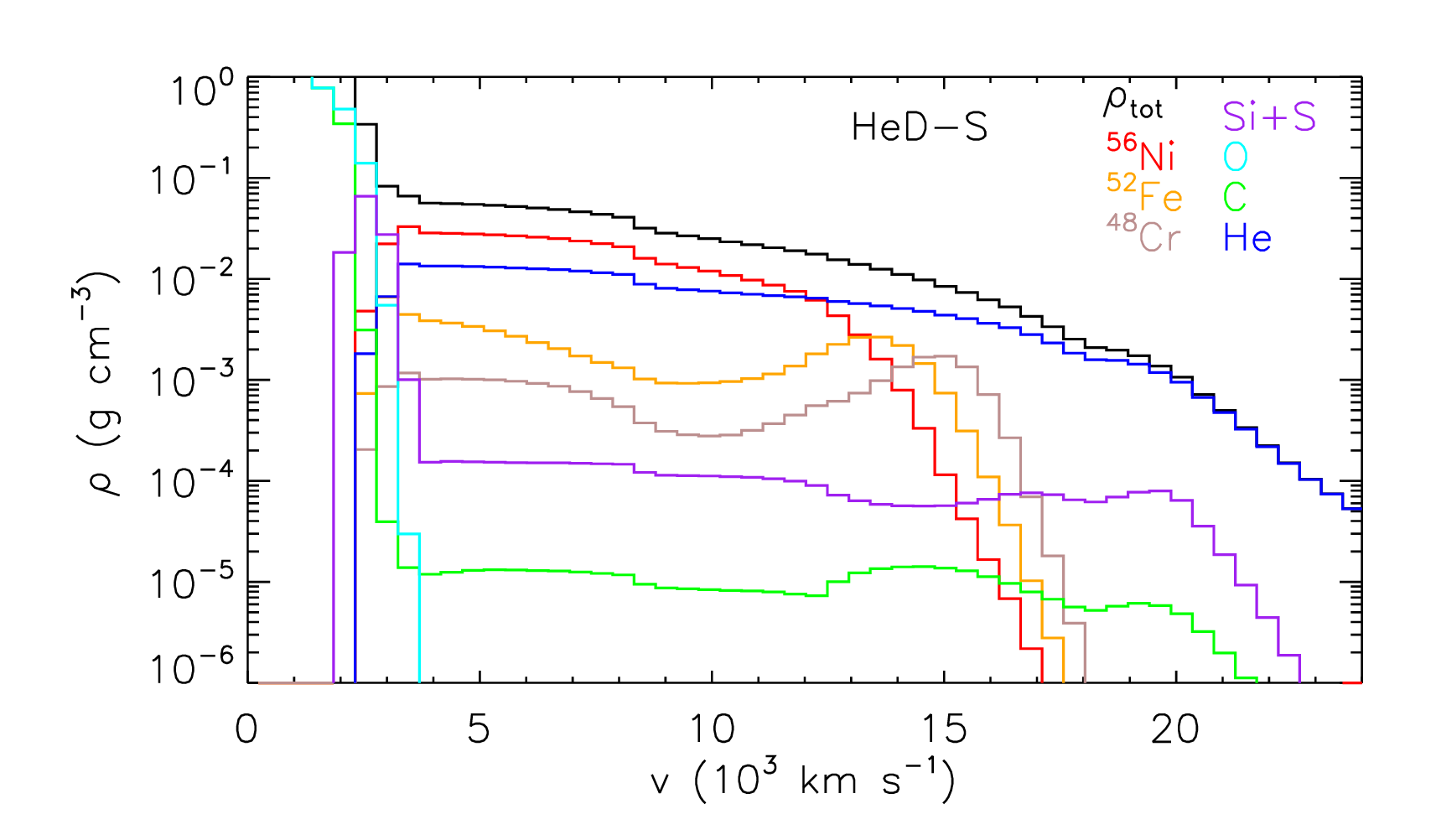,angle=90,width=9cm}
\caption{Composition of the explosion simulations for our S-model (CSDD-S, ELDD-S, HeD-S models, top to bottom).
The left panels show the mean atomic number ($\bar{Z}$) for each of the Lagrangian nucleosynthesis tracer particles at their final positions at the end of the hydrodynamical simulations (100~s after ignition of the He detonation). The models are symmetric under rotation about the $z$-axis. Most of the dense CO core material in the HeD-S model (the encircled dark blue region around the origin) remains bound and is not included in our radiative transfer simulations of the homologous ejecta. 
The right panels show the detailed composition for an equatorial slice through our simulations extrapolated to the homologous phase (the bound material in the HeD-S model is not included here). The black histograms show the total mass density ($\rho_{\mbox{\scriptsize tot}}$) versus
  expansion velocity. The contributions from various important isotopes and elements are indicated by the coloured histograms in each plot; specifically, they show $\rho_{\mbox{\scriptsize tot}} X_{i}$, where $X_{i}$ is the mass fraction of the species in question.}
\label{fig:compositions}
\end{figure*}

The mass yields obtained from the nucleosynthesis post-processing of
the CSDD models are tabulated in Table~\ref{tab:yields}. The
ejecta composition of model CSDD-S 
is illustrated in the top panels of Figure~\ref{fig:compositions}, which shows both the 2D distribution of mean atomic number and the detailed composition for a slice through the equatorial plane of the model.

\begin{table*}
\caption{Masses of important elements and isotopes in the ejecta for our converging-shock
  double-detonation (CSDD), edge-lit double-detonation (ELDD) and He-only
  detonation (HeD) simulations for our standard (S) and low-mass (L) models. The upper part of the table lists the masses of material originating in the outer He layer while the lower part refers to the products of the CO core. Only the unbound core material is included for the HeD models, having a total mass of 0.041 and 0.029~M$_{\odot}$ for models HeD-S and HeD-L, respectively.}
\begin{tabular}{llllllll}
  \hline
 &  & CSDD-S & ELDD-S & HeD-S & CSDD-L & ELDD-L & HeD-L\\ \hline
{He-layer ejecta} &
$^{56}$Ni (M$_{\odot}$) & $6.5 \times 10^{-2}$ & $7.2 \times 10^{-2}$ &  $6.5 \times 10^{-2}$  & $2.8 \times 10^{-3}$ &$7.6 \times 10^{-3}$& $2.8 \times 10^{-3}$\\
{composition}&$^{52}$Fe (M$_{\odot}$) & $1.4 \times 10^{-2}$ & $1.2 \times 10^{-2}$ & $1.4 \times 10^{-2}$ & $9.4 \times 10^{-3}$ &$1.4 \times 10^{-2}$& $9.4 \times 10^{-3}$\\
&$^{48}$Cr (M$_{\odot}$) & $8.6 \times 10^{-3}$ & $7.7 \times 10^{-3}$ & $8.6 \times 10^{-3}$ & $1.8 \times 10^{-2}$ &$1.9 \times 10^{-2}$& $1.8 \times 10^{-2}$\\
&Ti (M$_{\odot}$) & $3.7 \times 10^{-3}$ & $3.3 \times 10^{-3}$ & $3.7 \times 10^{-3}$ &  $1.1 \times 10^{-2}$ &$9.6 \times 10^{-3}$&  $1.1 \times 10^{-2}$\\
&Ca (M$_{\odot}$) & $8.8 \times 10^{-3}$ & $8.2 \times 10^{-3}$ & $8.8 \times 10^{-3}$ & $2.7 \times 10^{-2}$ &$2.5 \times 10^{-2}$& $2.7 \times 10^{-2}$ \\
&S  (M$_{\odot}$) & $1.7 \times 10^{-3}$ & $2.5 \times 10^{-3}$ & $1.7 \times 10^{-3}$ & $6.0 \times 10^{-3}$ &$6.5 \times 10^{-3}$& $6.0 \times 10^{-3}$\\
&Si (M$_{\odot}$) & $1.5 \times 10^{-3}$ & $6.0 \times 10^{-3}$ & $1.5 \times 10^{-3}$ & $1.9 \times 10^{-3}$&$4.9 \times 10^{-3}$& $1.8 \times 10^{-3}$\\
&Mg (M$_{\odot}$) & $7.4 \times 10^{-4}$ & $8.1 \times 10^{-4}$ & $7.4 \times 10^{-4}$ & $5.9 \times 10^{-4}$ &$1.3 \times 10^{-3}$& $5.9 \times 10^{-4}$\\
&O (M$_{\odot}$) & $1.0 \times 10^{-3}$ & $1.4 \times 10^{-3}$ & $1.0 \times 10^{-3}$ & $6.9 \times 10^{-4}$ &$1.4 \times 10^{-3}$& $6.9 \times 10^{-4}$\\
&C (M$_{\odot}$) & $1.9 \times 10^{-4}$ & $1.5 \times 10^{-4}$ & $1.8 \times 10^{-4}$ & $1.5 \times 10^{-3}$ &$9.4 \times 10^{-4}$& $1.5 \times 10^{-3}$ \\
&He (M$_{\odot}$) & $8.0 \times 10^{-2}$ & $7.3 \times 10^{-2}$ & $8.0 \times 10^{-2}$ & $1.2 \times 10^{-1}$&$1.1 \times 10^{-1}$& $1.2 \times 10^{-1}$\\
\hline
{CO core ejecta} &
$^{56}$Ni (M$_{\odot}$) & $1.5 \times 10^{-1}$ &  $5.6 \times 10^{-3}$ & -- & $1.9 \times 10^{-2}$&$1.5 \times 10^{-7}$& --\\
{composition}&$^{52}$Fe (M$_{\odot}$) & $4.0 \times 10^{-3}$ &  $4.5 \times 10^{-4}$ & -- & $9.3 \times 10^{-4}$&$4.5 \times 10^{-9}$& --\\
&$^{48}$Cr (M$_{\odot}$) & $2.0 \times 10^{-4}$ &  $4.4 \times 10^{-5}$ & -- & $5.8 \times 10^{-5}$&$5.7 \times 10^{-9}$& --\\
&Ti (M$_{\odot}$) & $6.8 \times 10^{-6}$ &  $3.7 \times 10^{-6}$ & -- & $2.8 \times 10^{-6}$ &$3.8 \times 10^{-8}$ & --\\
&Ca (M$_{\odot}$) & $1.1 \times 10^{-2}$ &  $1.0 \times 10^{-2}$ &$3.2 \times 10^{-9}$  & $6.2 \times 10^{-3}$ &$1.7 \times 10^{-4}$& --\\
&S  (M$_{\odot}$) & $7.5 \times 10^{-2}$ &  $1.0 \times 10^{-1}$ &$2.3 \times 10^{-5}$  & $5.4 \times 10^{-2}$ &$1.1 \times 10^{-2}$& $3.5 \times 10^{-9}$\\
&Si (M$_{\odot}$) & $1.8 \times 10^{-1}$ &  $2.9 \times 10^{-1}$ &$6.1 \times 10^{-4}$  & $1.6 \times 10^{-1}$&$1.2 \times 10^{-1}$& $7.9 \times 10^{-7}$\\
&Mg (M$_{\odot}$) & $2.8 \times 10^{-2}$ &  $2.5 \times 10^{-2}$ & $1.6 \times 10^{-3}$ & $4.0 \times 10^{-2}$ &$7.6 \times 10^{-2}$& $1.2 \times 10^{-5}$\\
&O (M$_{\odot}$) & $9.3 \times 10^{-2}$ &  $1.1 \times 10^{-1}$ &$2.0 \times 10^{-2}$  & $1.3 \times 10^{-1}$ &$2.3 \times 10^{-1}$& $1.4 \times 10^{-2}$\\
&C (M$_{\odot}$) & $1.6 \times 10^{-2}$ &  $9.0 \times 10^{-3}$ & $1.7 \times 10^{-2}$ & $2.2 \times 10^{-2}$ &$1.3 \times 10^{-2}$& $1.4 \times 10^{-2}$\\
&He (M$_{\odot}$) & $3.5 \times 10^{-4}$ & -- & -- & $1.9 \times 10^{-5}$& -- & --\\
\hline
\end{tabular}
\label{tab:yields}
\end{table*}

For our CSDD-S model,
a significant mass of radioactive nuclei (specifically ${\sim}0.08$~M$_{\odot}$ of $^{56}$Ni and $^{52}$Fe) is produced by the detonation of the He layer. This yield of radioactive nuclei is similar to that found by \citet{shen10} and \citet{waldman11} for models with comparable values of $\rho_\text{b}$ (specifically, our pattern of radioactive yields lies between those of \citealt{shen10} for detonations of 0.2 and 0.3~M$_{\odot}$ He-layers around 0.6~M$_{\odot}$ CO cores). 
The decay of this material will power the early phases of the light curve and produce a transient that brightens rapidly, on a timescale 
of several days, as predicted by \citet{bildsten07} (see Section~\ref{sect:obs}). In our model, however, the shock compression is sufficient to yield an even larger mass of $^{56}$Ni (${\sim}0.1$~M$_{\odot}$) from the core detonation. As in double-detonation models for more massive CO WDs {\citep[e.g.][]{woosley94,livne95a,fink07,woosley11}},
this $^{56}$Ni is concentrated at low velocities and surrounded by an envelope of intermediate mass elements, predominantly silicon and sulphur
(see Figure~\ref{fig:compositions}).
Since this centrally concentrated $^{56}$Ni is enshrouded by a much larger mass envelope than the $^{56}$Ni from the He-detonation, the outwards diffusion time will be longer. Thus the light curve will evolve on longer timescales than for the p-Ia events predicted by \citet{bildsten07}, \citet{shen09} and \citet{shen10}; see Section~\ref{sect:obs}. 

Qualitatively similar results are found for our CSDD-L model. In this case, $^{48}$Cr is the dominant radioactive product from burning of the He layer. Also, the low density of the CO core means that the $^{56}$Ni mass produced in the core detonation is now smaller than the mass of radioactive elements produced in the He-shell. Nevertheless, the core $^{56}$Ni mass is not negligible and affects the synthetic observables, as will be discussed in Section~\ref{sect:obs}.

\section{Additional models}
\label{sect:othermods}

For comparison of the results obtained with our CSDD models, we have performed additional simulations to quantify the observable properties of alternative explosion mechanisms. These were set up and carried out in an identical manner except that the detonation of the CO core was handled differently.

\subsection{He detonation only; HeD models}

Although the simulations described in Section~\ref{sect:csdd} suggest that detonation of the surrounding He layer will trigger a secondary detonation of the core, the difficulty in determining whether a detonation is initiated must be recognized \citep[e.g.][see also Section~\ref{sect:core_det}]{seitenzahl09a}.
Therefore, as an experiment, we also performed calculations 
in which it is assumed that no core detonation is ignited
(hereafter, our HeD models, which were carried out for both our S and L initial systems). 
These models are an important comparison point
since they are the realisations of the \citet{bildsten07} p-Ia explosion
scenario that correspond to our CSDD models. 

For these simulations, the He detonation was ignited exactly as before
and produces very similar nucleosynthetic yields to the He layer in
the CSDD model (Table~\ref{tab:yields}). Since it is assumed that no
core detonation takes place, most of the underlying CO core is
unaffected by the He detonation and remains tightly bound\footnote{We
note, however, that even in this case the heating in the shock
convergence leads to burning a
small fraction of the mass to $^{56}$Ni (${\sim}10^{-4}$~M$_{\odot}$ for
our S-model) and intermediate-mass elements (${\sim}10^{-3}$~M$_\odot$).}.
This bound material is not included in our radiative transfer simulations, which involve only the homologously expanding ejecta.
However, part of the CO core (${\sim}0.041$~M$_{\odot}$ for HeD-S and 0.029~M$_{\odot}$ for HeD-L)
is unbound as a result of kinetic energy transferred from the He
detonation to material of the CO core \citep[see also][]{woosley86}. Therefore, some material from the core is still present in the ejecta and dominates the composition at low velocities (see lower right panel of Figure~\ref{fig:compositions}).

\subsection{Edge-lit core detonation; ELDD}
\label{sect:ELDD}

An alternative to the CSDD model is that the He detonation
directly ignites 
an inward propagating detonation at the edge of the CO
core. 
Whether such a detonation can be ignited depends on many
factors including the density at the edge of the CO core ($\rho_\text{b}$), the
composition and the geometry. 1D simulations have suggested that it is
most likely to happen if the ignition point is some way
above the base of the He-shell \citep[see
e.g.][]{nomoto82,livne90a,benz97a,garcia99a}. Edge-lit detonation has
been found in some multi-dimensional simulations \cite[e.g.][]{livne91a}
although it may be harder when the 1D symmetry is broken \citep{forcada07a}.

Edge-lit 
CO detonation requires that densities of at least $10^6$~g~cm$^{-3}$ and critical temperatures of several billion Kelvin are reached in the outer CO material \citep{roepke07}. In the simulations for our S-model, we do find that some regions at the very edge of the CO core are heated to temperatures $\simgt 2 \times 10^9$~K and that densities in excess of $10^6$~g~cm$^{-3}$ are reached in some places (particularly close to the He detonation convergence point on the $-z$-axis). However, these hot/dense conditions appear in only a handful of our nucleosynthesis tracer particles and it is therefore marginal whether critical volumes for detonation are really reached. Moreover, since our models are not based on evolutionary calculations, we cannot 
predict at what height in the He-layer ignition of the detonation
is most likely to occur -- we have simply chosen to ignite our He detonations at the base of our He layers.
Thus, our simulations are ill-suited to determine whether edge-lit detonation is probable in our particular systems.

Nevertheless, we can
investigate the observable consequences if such an edge-lit detonation
were to occur -- to do this we
performed simulations in which a CO detonation was ignited by hand at the edge
of the core immediately below the ignition point of the He detonation.
The two detonations were ignited simultaneously but modelled
independently.
We will refer to the results of these simulation as our edge-lit double-detonation models (ELDD-S and ELDD-L, for our two initial systems, respectively). 
The important difference from the CSDD models is that there is no
strong shock convergence in the core prior to ignition of the CO
detonation. This means that the densities in the core remain close to
their initial values, which are too low to lead to significant
$^{56}$Ni production in the core. Thus, although sufficient energy
is released to unbind the core, the core is primarily burned to
intermediate mass elements
(see Table~\ref{tab:yields} and Figure~\ref{fig:compositions}). 
In addition, these models have slightly more complete burning of the
He shell material, a consequence of additional heating of the
burning region just behind the He detonation by oblique shocks generated
from the CO detonation.

\section{Synthetic observables}
\label{sect:obs}

\subsection{Light curve morphology}

\begin{figure*}
\epsfig{file=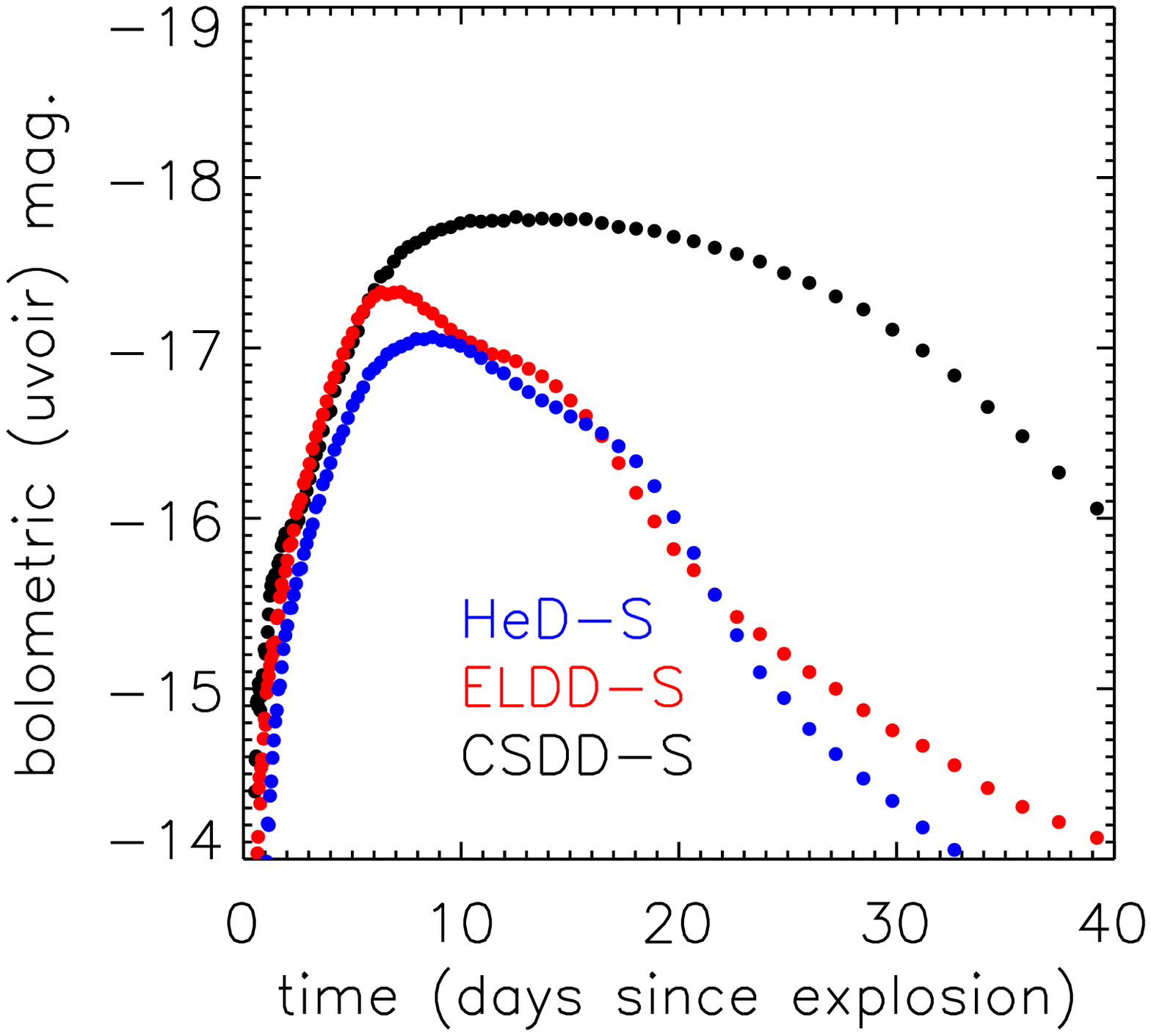,angle=90,width=5cm}
\epsfig{file=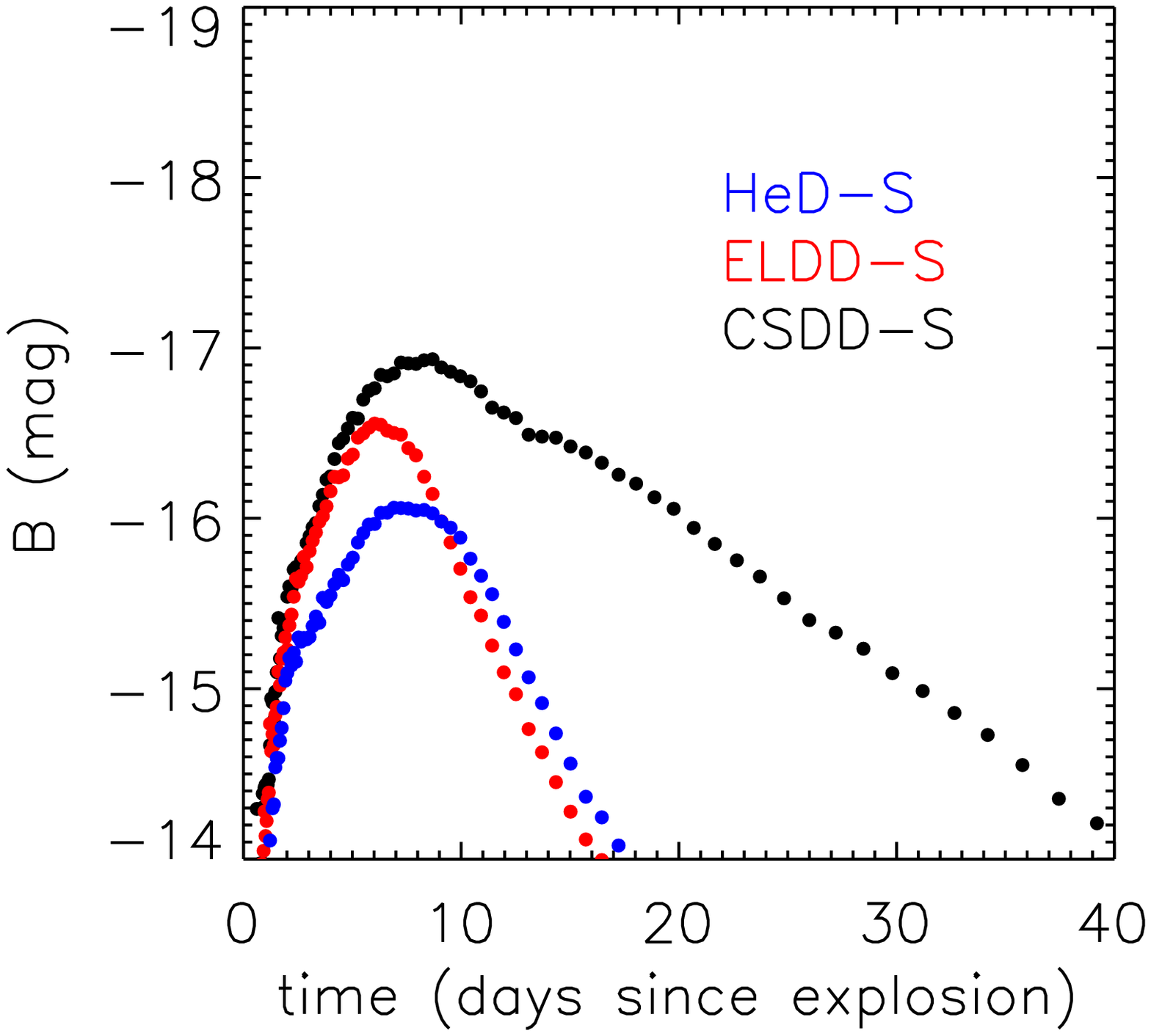,angle=90,width=5cm}
\epsfig{file=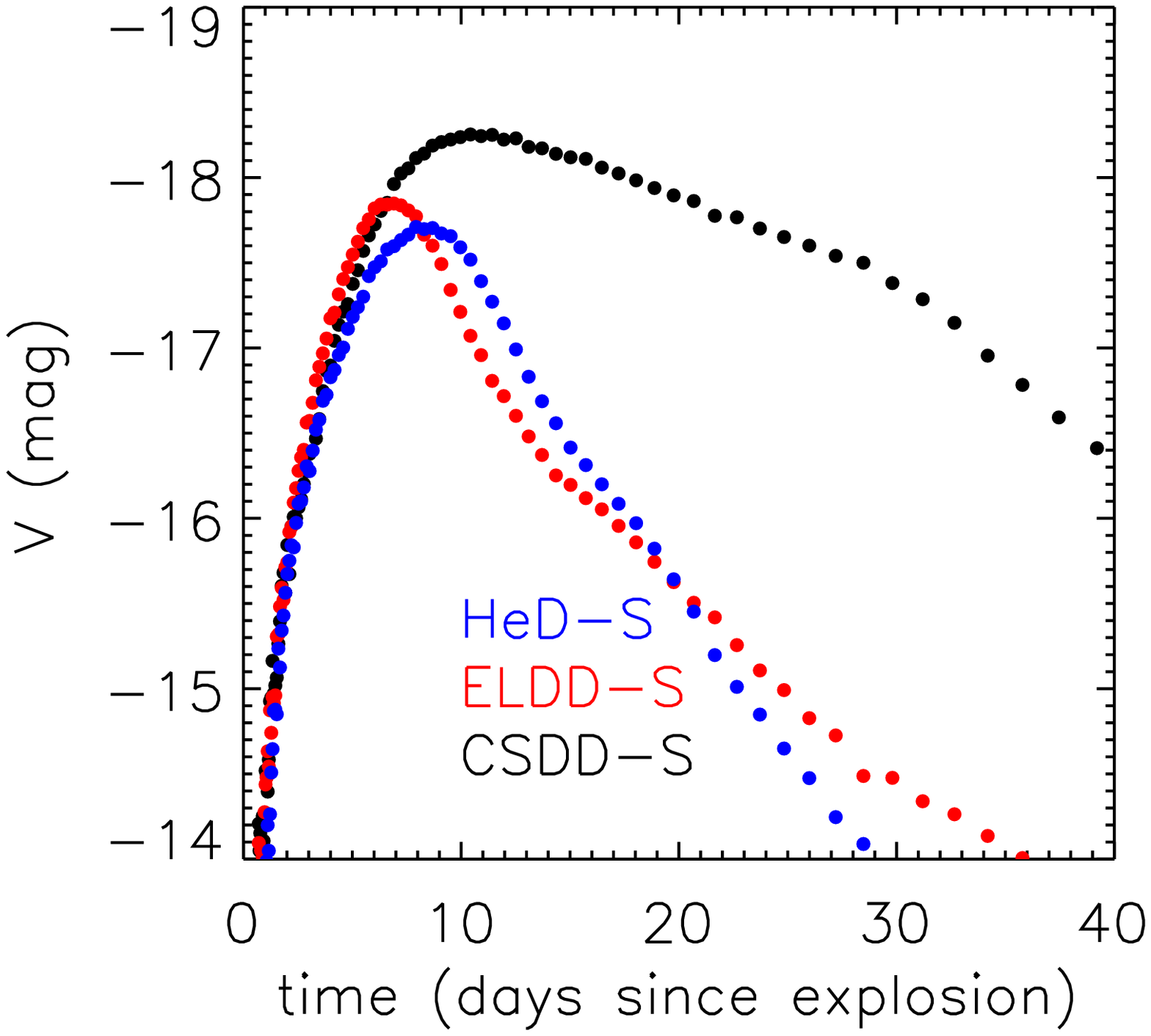,angle=90,width=5cm}\\
\epsfig{file=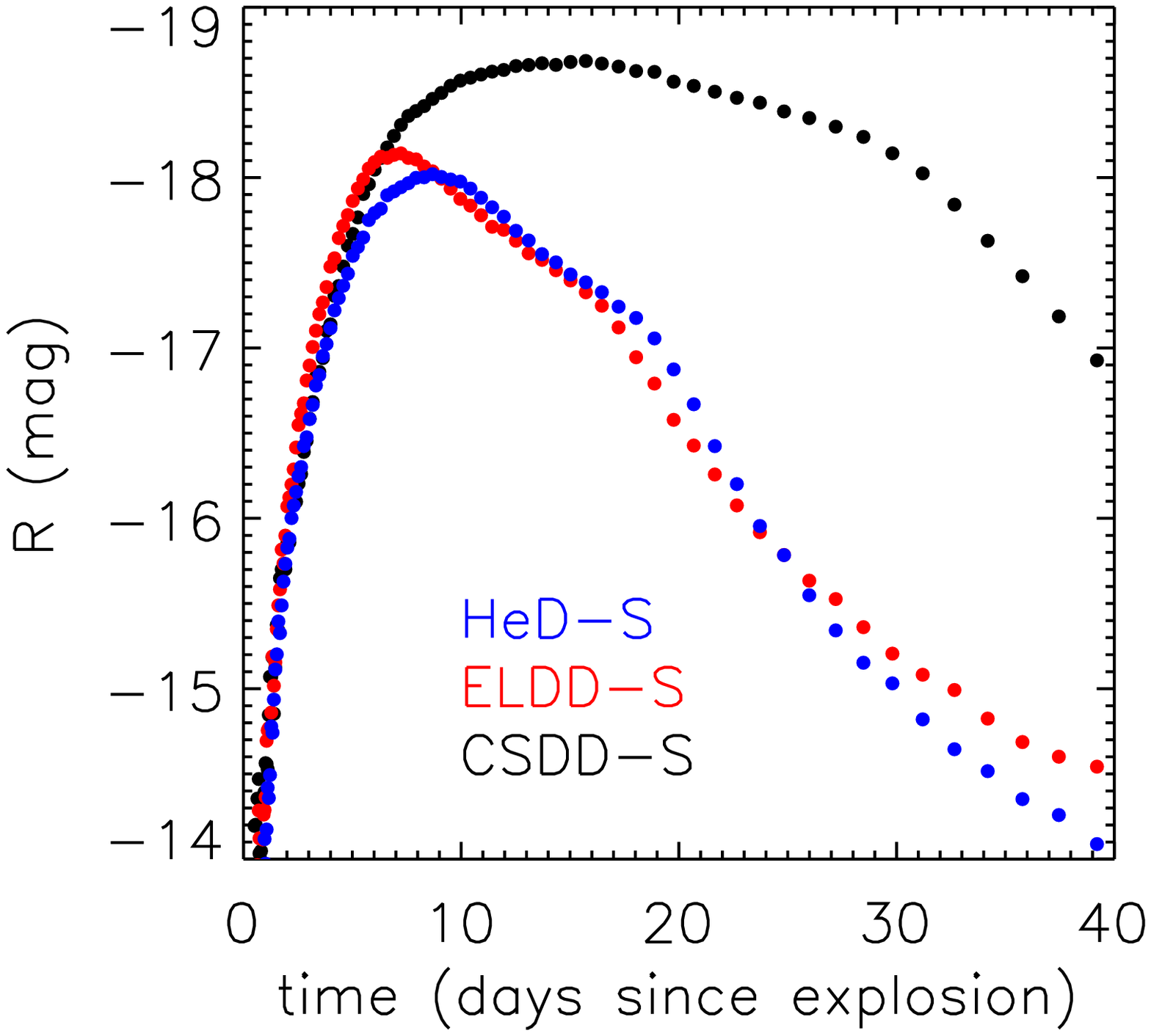,angle=90,width=5cm}
\epsfig{file=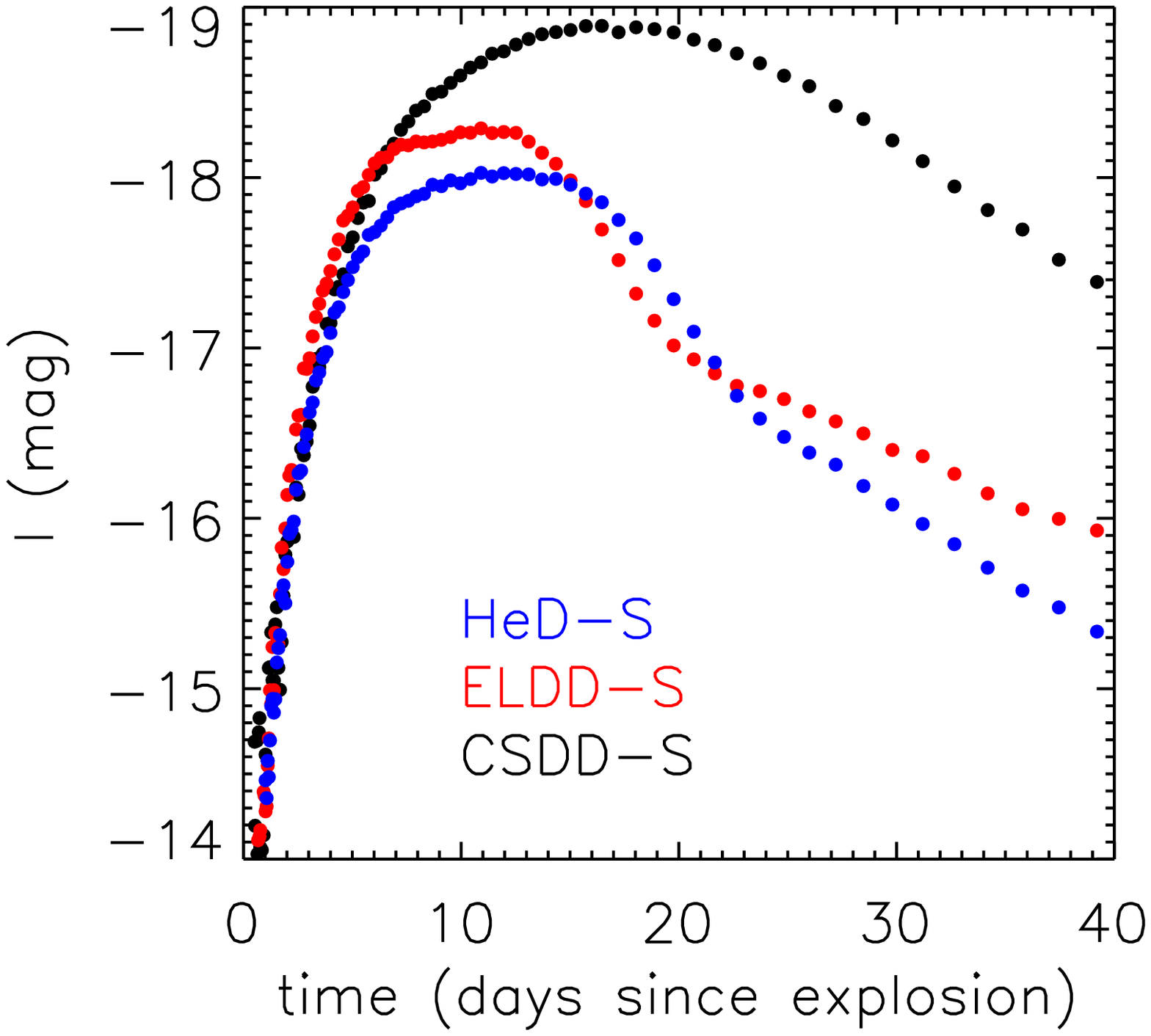,angle=90,width=5cm}
\epsfig{file=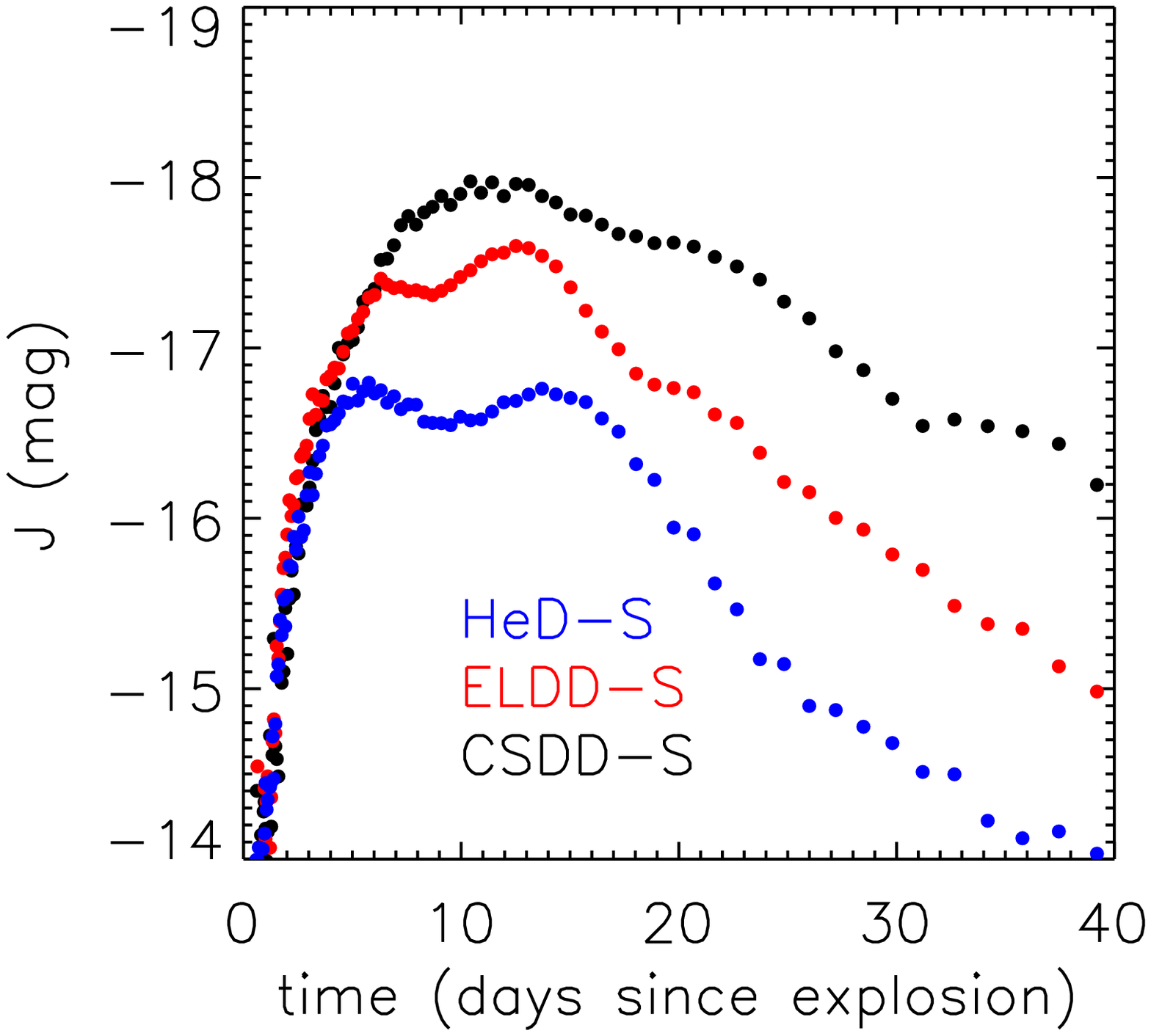,angle=90,width=5cm}\\
\caption{Angle-averaged light curves for our explosion simulations based on our S-model (CSDD-S, ELDD-S and HeD-S) in bolometric (ultraviolet--optical--infrared, \textit{UVOIR}), $B$-, $V$-, $R$-, $I$- and $J$-bands.}
\label{fig:lcs}
\end{figure*}

\begin{figure*}
\epsfig{file=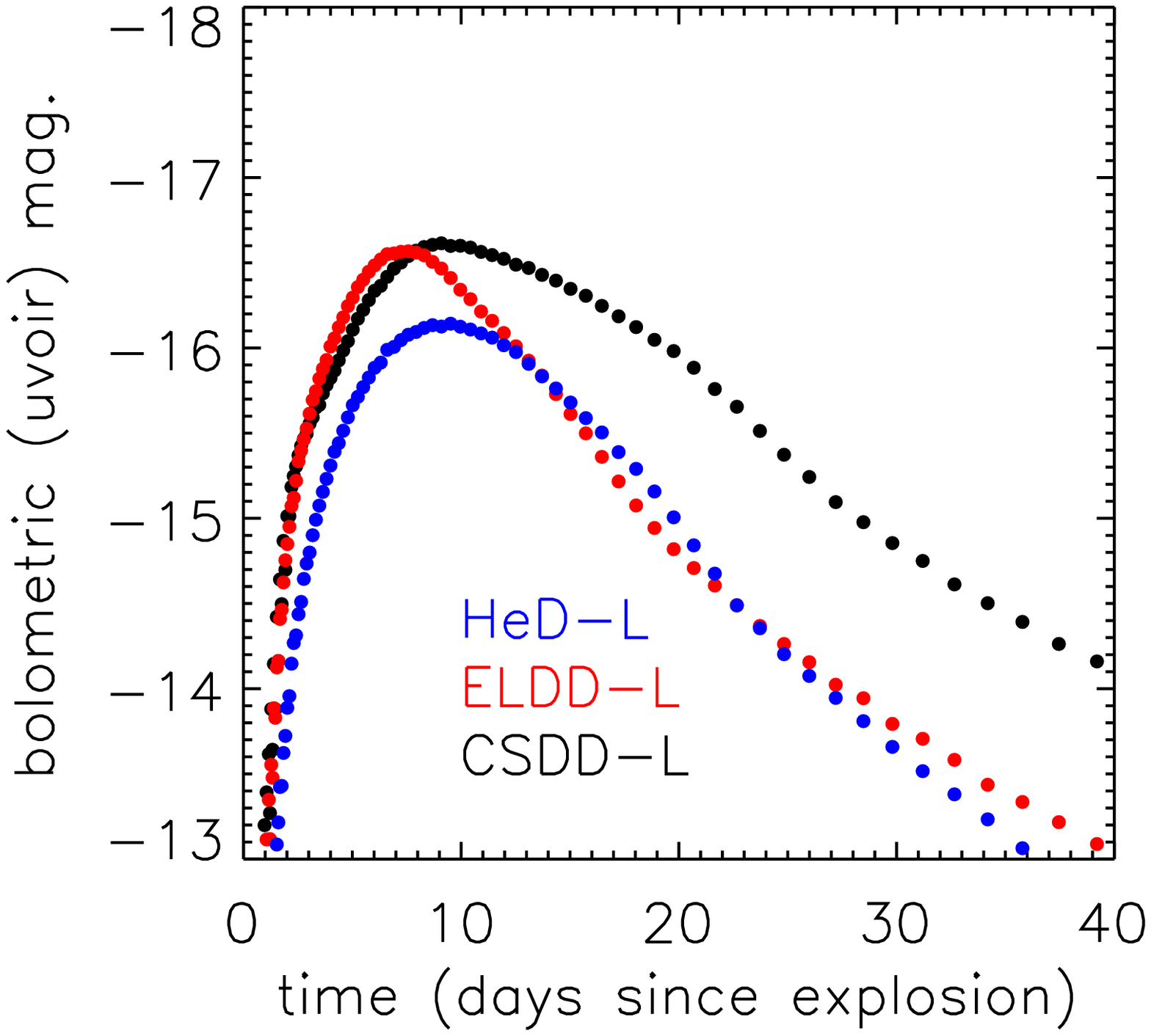,angle=90,width=5cm}
\epsfig{file=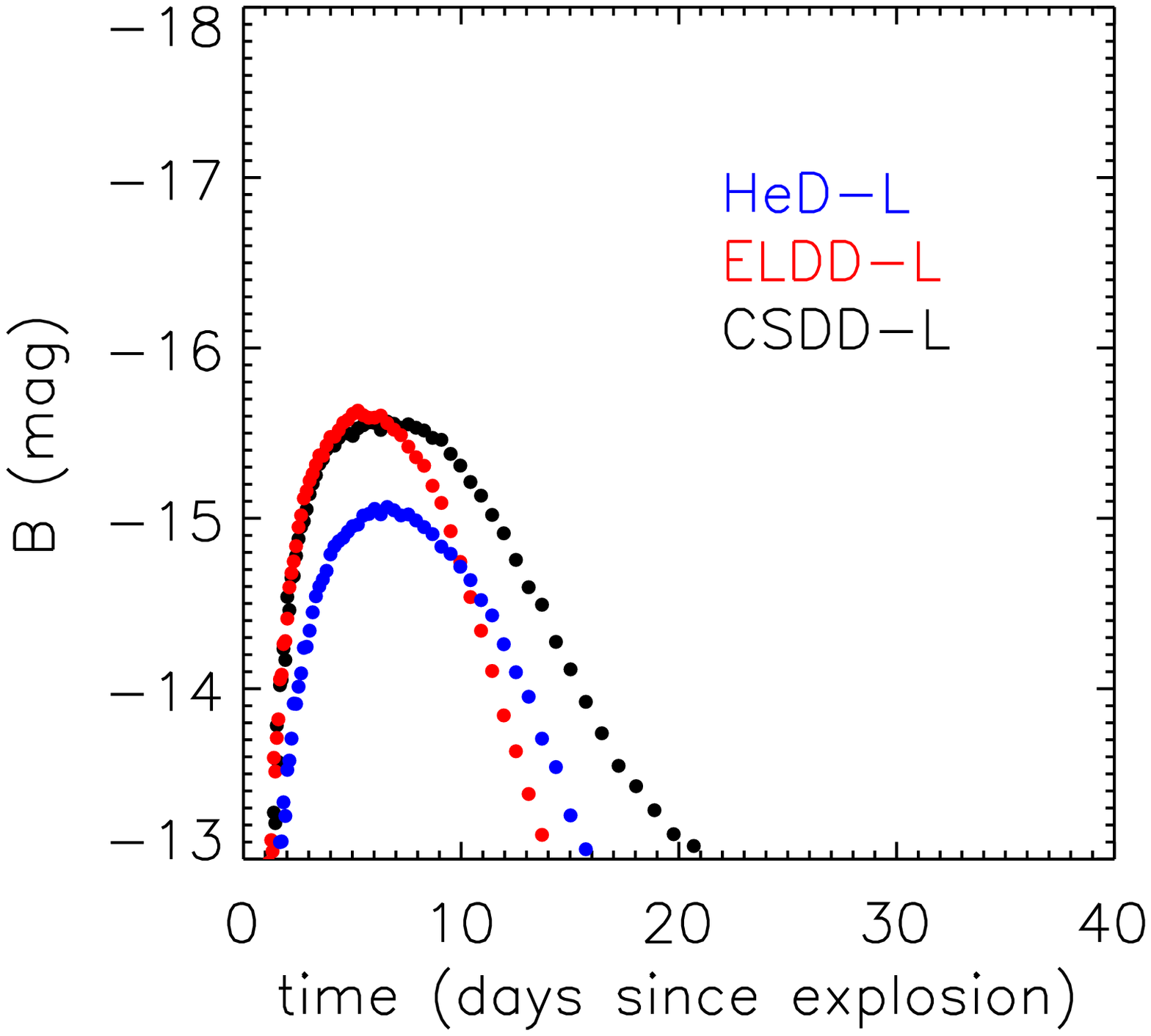,angle=90,width=5cm}
\epsfig{file=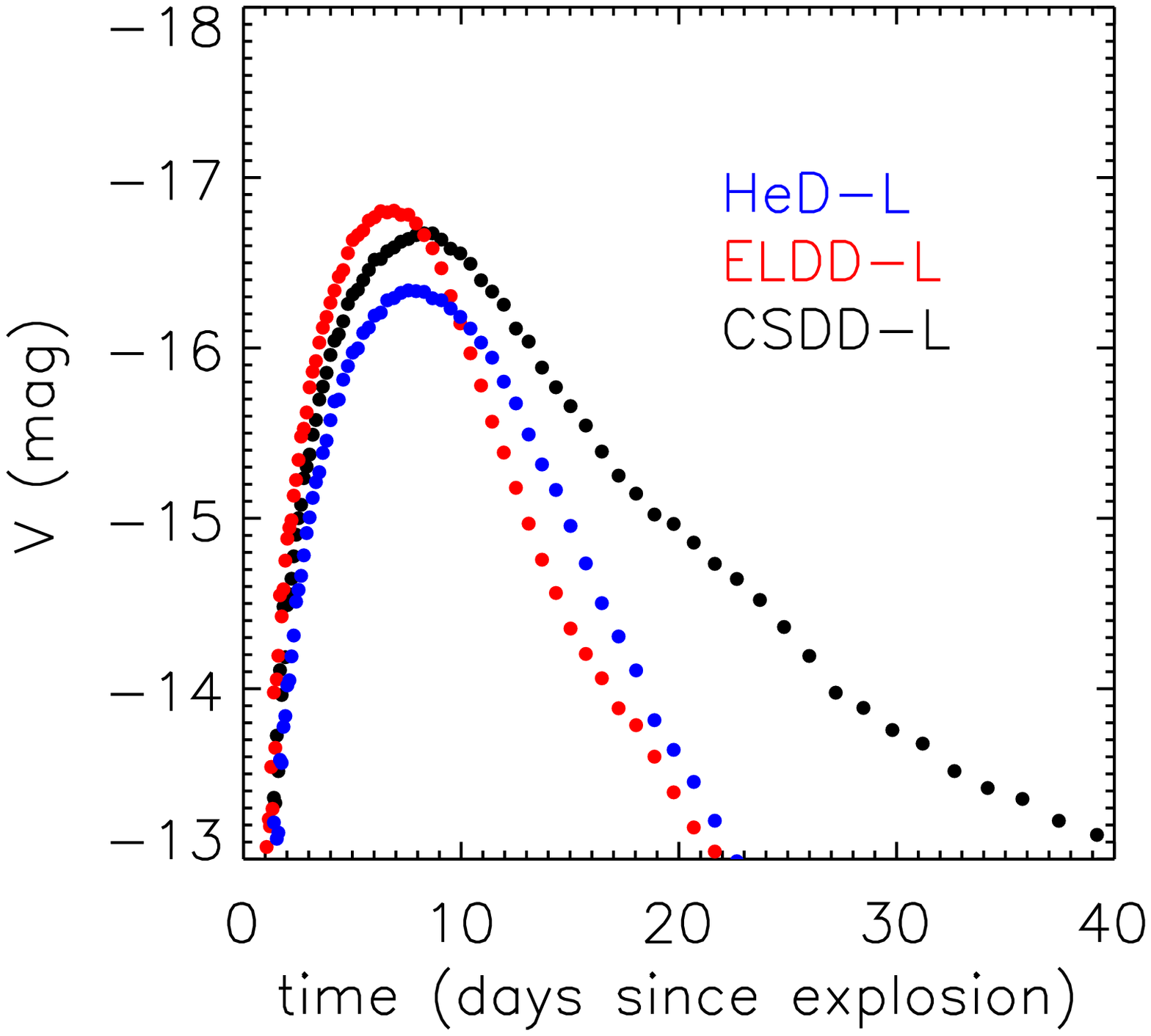,angle=90,width=5cm}\\
\epsfig{file=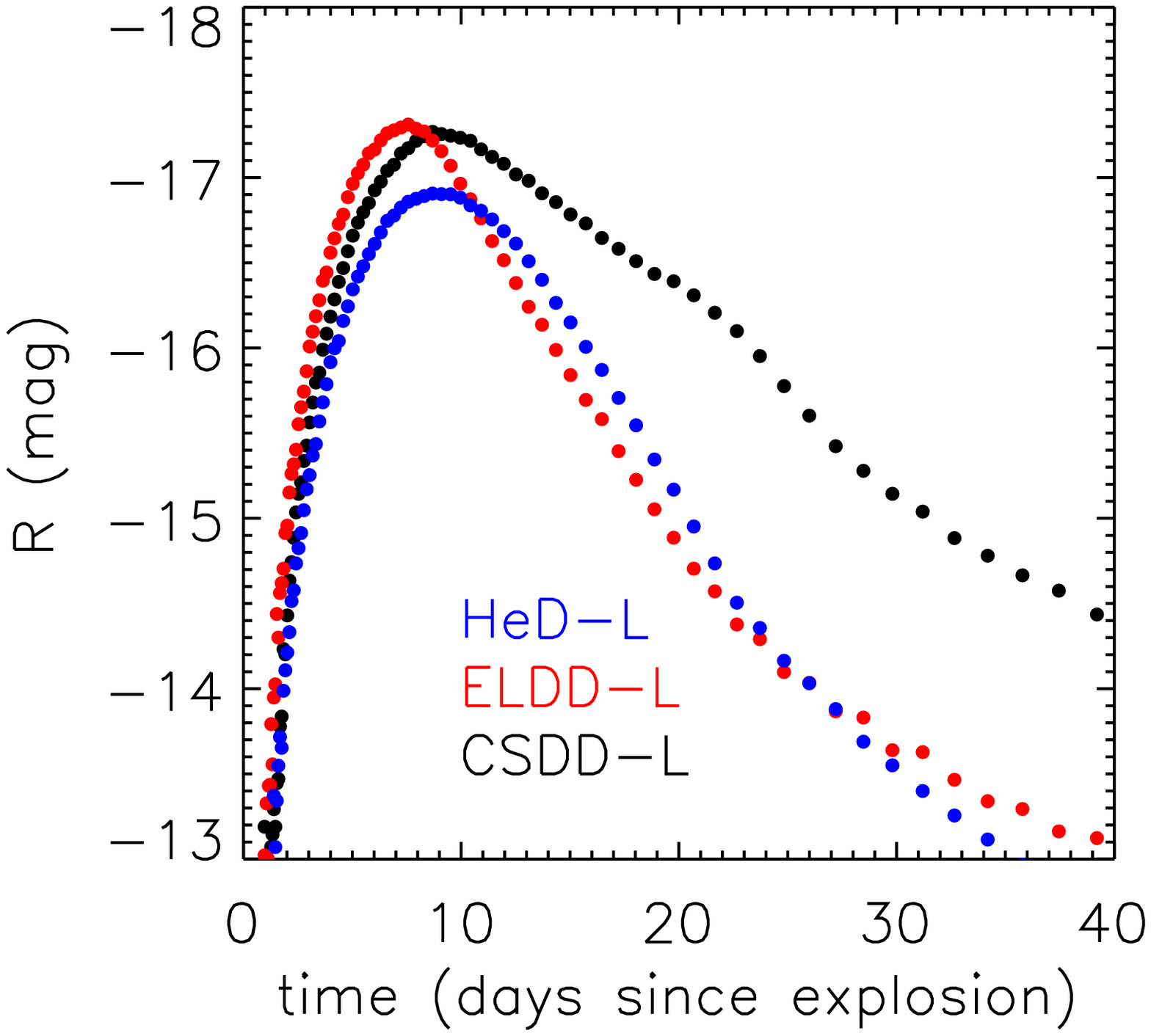,angle=90,width=5cm}
\epsfig{file=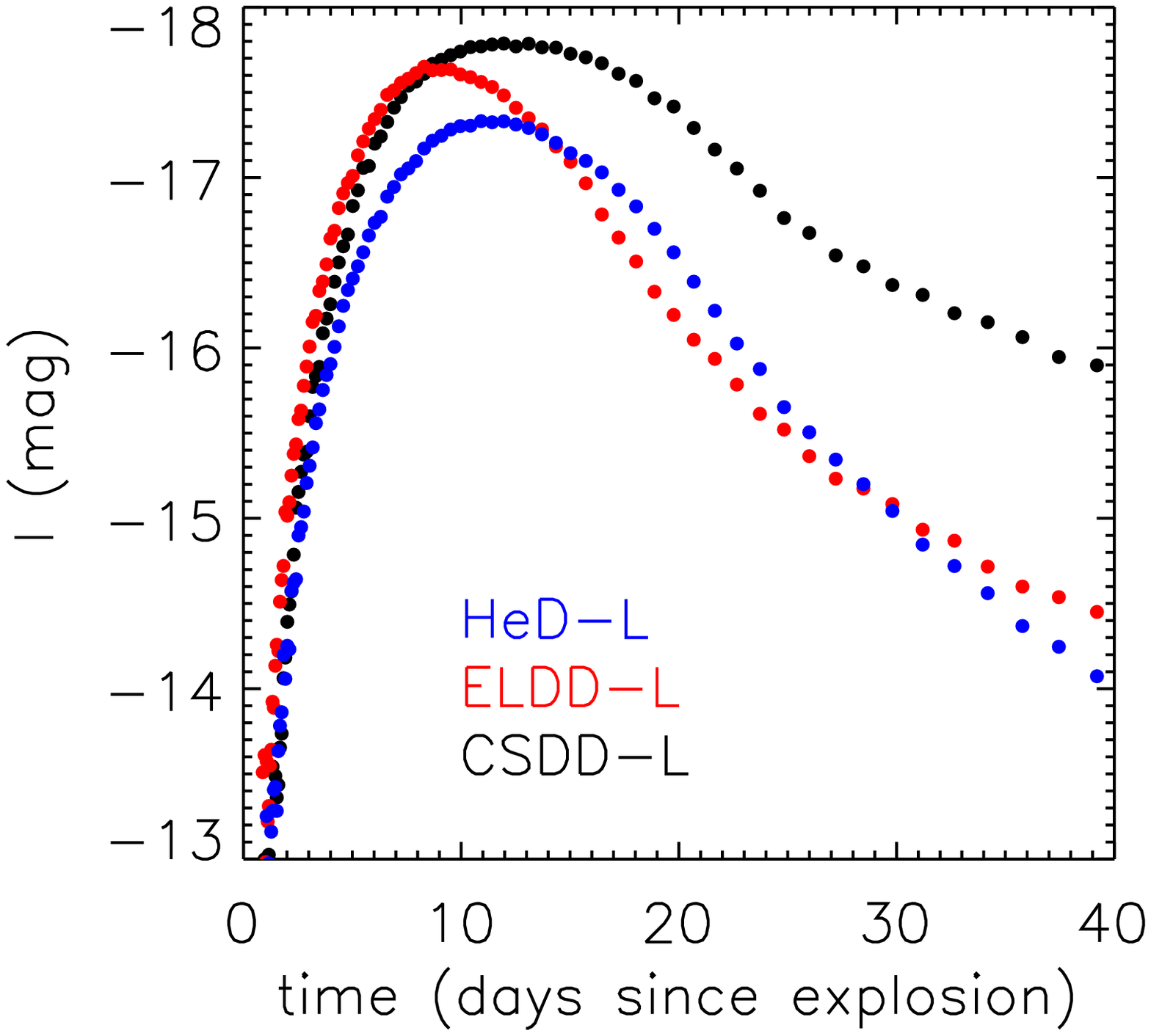,angle=90,width=5cm}
\epsfig{file=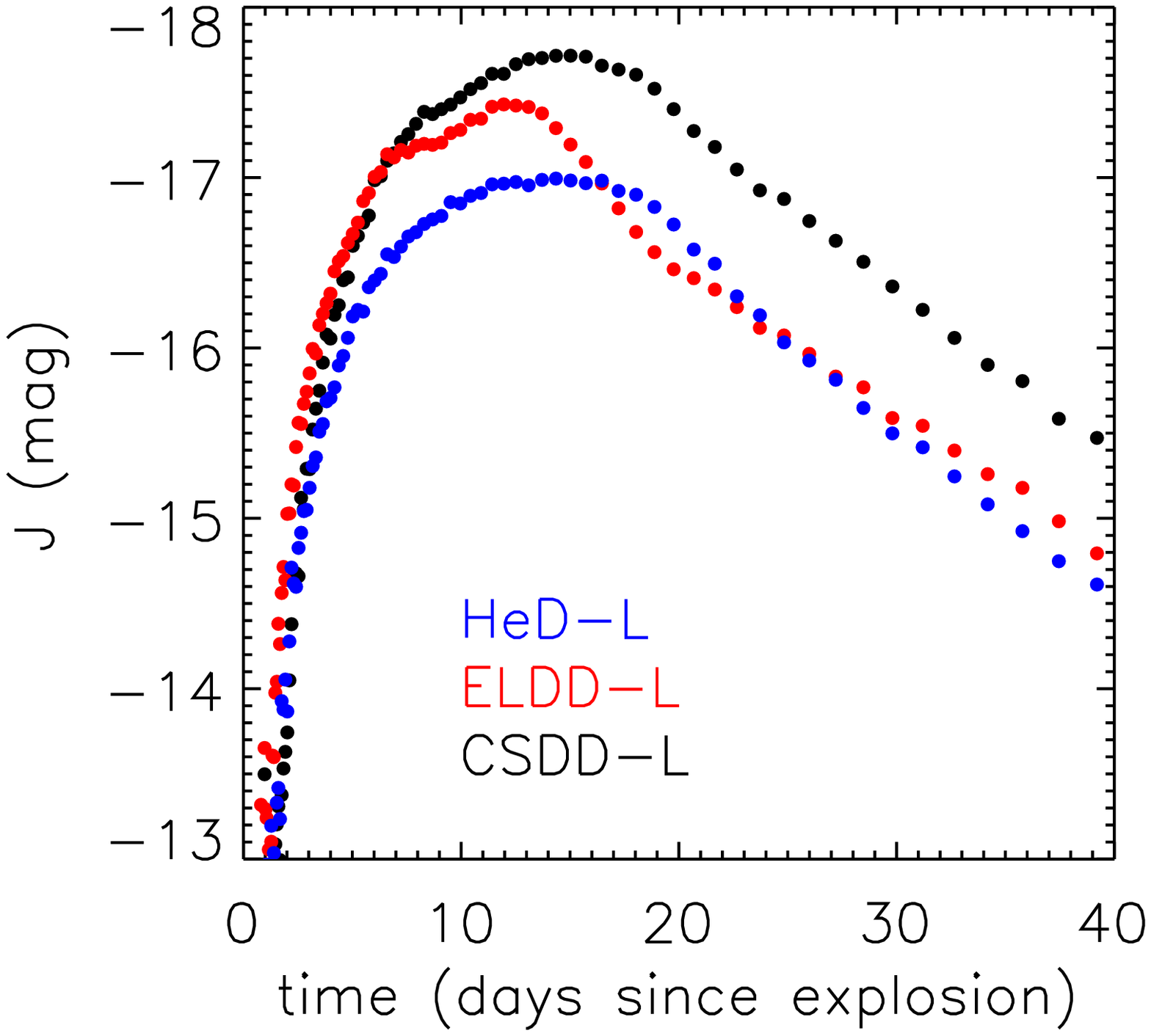,angle=90,width=5cm}\\
\caption{As Figure~\ref{fig:lcs} but showing results for simulations with our L-model (CSDD-L, ELDD-L and HeD-L).}
\label{fig:lcsL}
\end{figure*}

Figures~\ref{fig:lcs} and \ref{fig:lcsL} show synthetic bolometric (ultraviolet--optical--infrared; hereafter \textit{UVOIR}), optical- and infrared-band light curves computed for the three explosion mechanisms (CSDD, ELDD, HeD) for our two model systems (S and L, respectively).
Here, we show the angle-average synthetic light curves -- the dependence of the light curve properties on observer inclination will be discussed briefly in Section~\ref{sect:viewing_angle}.

\begin{figure*}
\epsfig{file=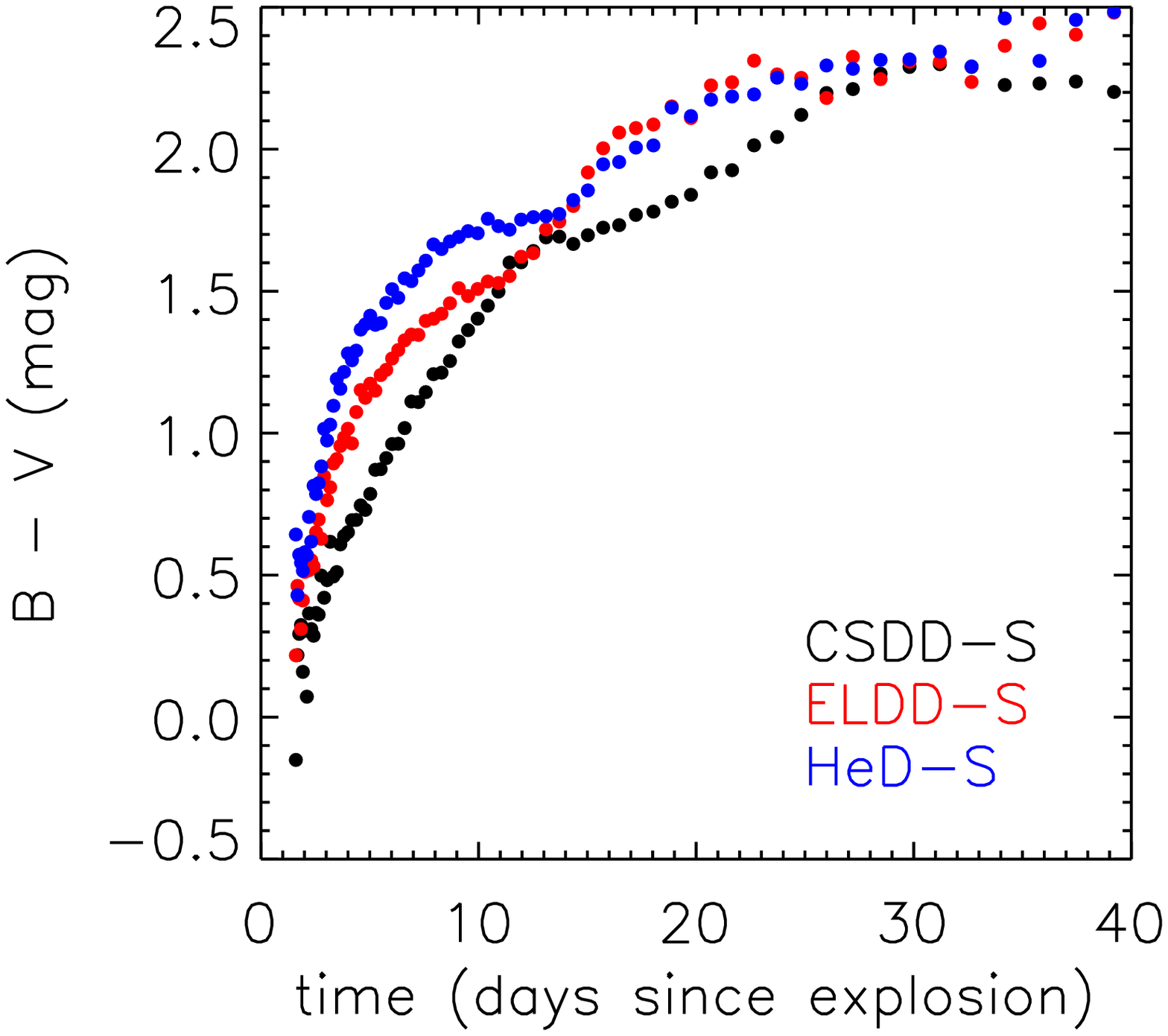,angle=90,width=5cm}
\epsfig{file=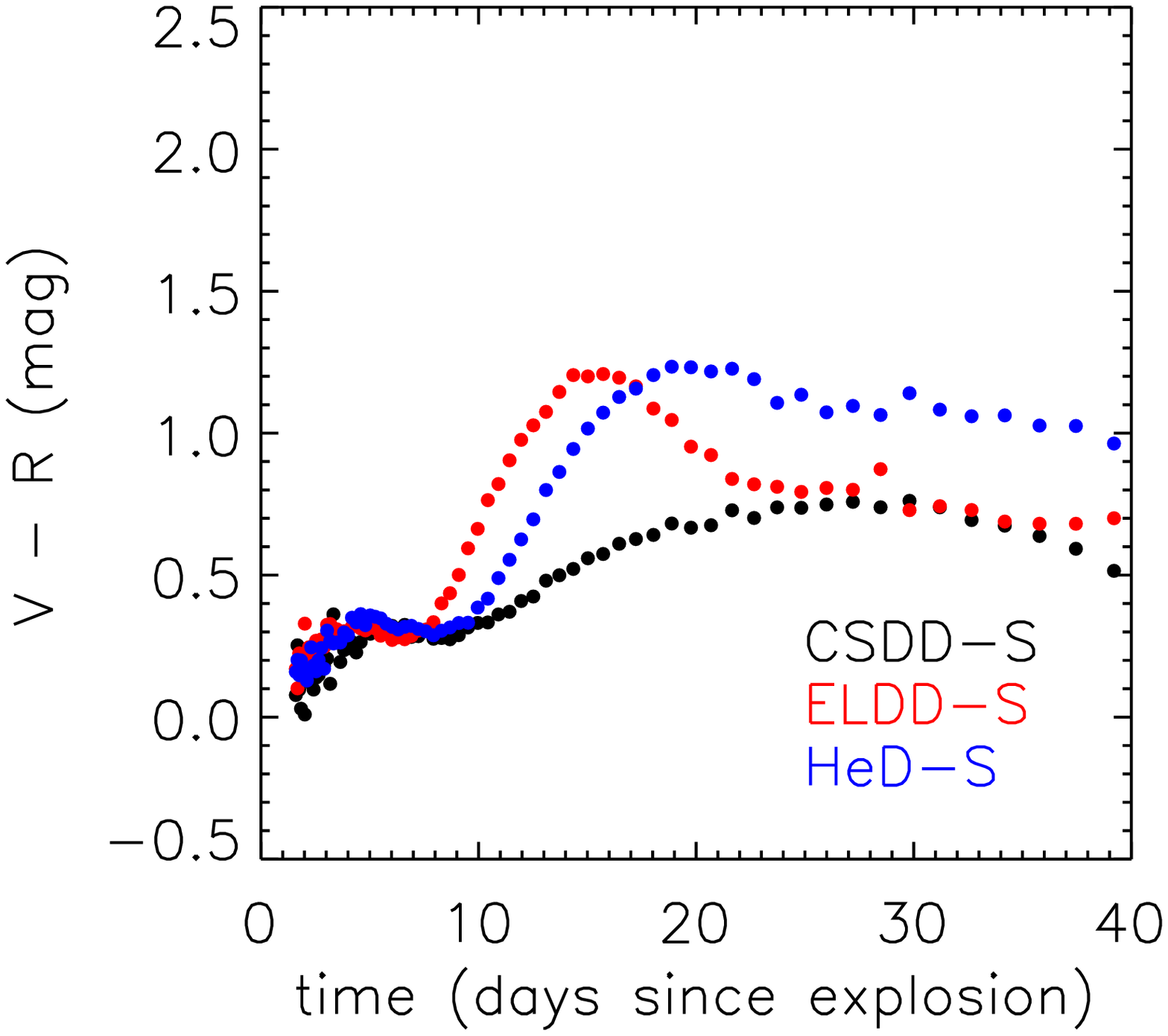,angle=90,width=5cm}
\epsfig{file=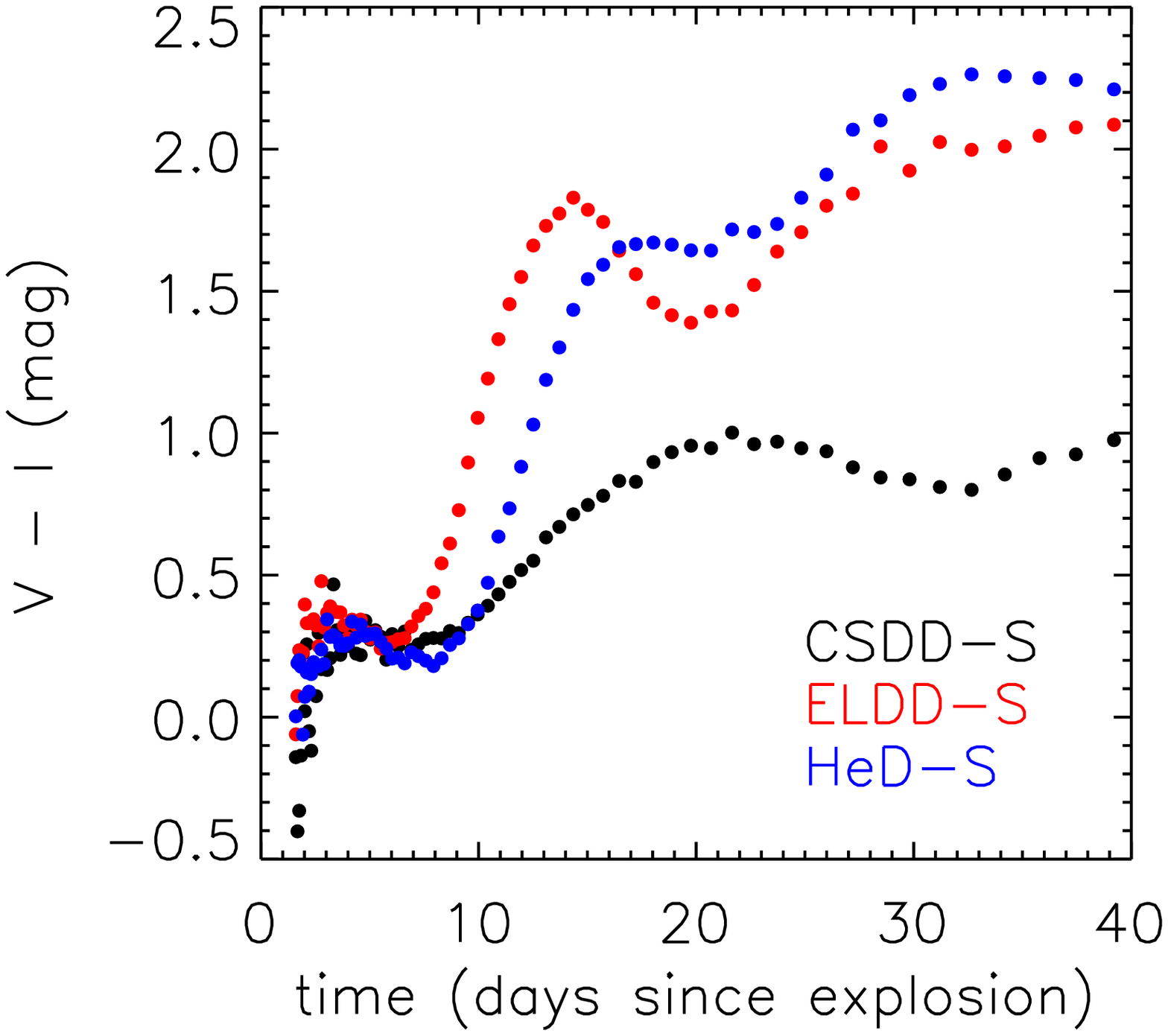,angle=90,width=5cm}\\
\caption{Optical colour evolution ($B - V$, $V - R$, $V - I$) for our three explosion simulations (CSDD-S, ELDD-S, HeD-S).}
\label{fig:colours}
\end{figure*}

It is apparent from Figure~\ref{fig:lcs} that the CSDD scenario leads to a very different transient from a p-Ia model (i.e.\ the HeD model) for our S-model. 
In agreement with the calculations for p-Ia models by \citet{shen10}, 
our HeD-S bolometric light curve reaches peak around 8~days after explosion and then decays fairly rapidly, dropping by ${\sim}1$~mag during a two-week period after maximum. In contrast, the CSDD-S model takes several days longer to reach peak and remains bright for an extended period -- significant bolometric decline does not commence until ${\sim}30$~days after explosion. The luminosity of the CSDD-S model is always higher than the HeD-S model.
Similar conclusions are drawn from our low-mass model (compare CSDD-L and HeD-L in Figure~\ref{fig:lcsL}). Here the scale of the effect is less extreme but the slower light curve evolution is still very apparent in the optical bands.

These differences can be understood as consequences of $^{56}$Ni-rich material produced in the core detonation (see Table~\ref{tab:yields}). Decay of $^{56}$Ni in the core produces a comparable amount of energy to that supplied by decay of radioactive nuclei in the outer layer of He-burning products. Moreover, the $^{56}$Ni in the core is deep inside the ejecta meaning that this energy takes longer to diffuse outward and $\gamma$-ray trapping is more effective for a longer period of time. This causes the slow light curve evolution. The scale of these effects is large and easily observable, corresponding to differences in excess of a magnitude in most bands at post-maximum epochs for our S-model.

The influence of the core material in the ELDD models is considerably
more subtle. 
From about 8~days after explosion, the ELDD-S bolometric
and optical light curves are much more similar to the HeD-S than CSDD-S
light curves. This is because the ELDD-S light curve is
predominantly powered by the radioactive nuclei produced in the He
detonation; the small mass of $^{56}$Ni in the core of the ELDD-S model
only becomes a noticeable energy source well after maximum light (at
times greater than ${\sim}20$~days after explosion, the ELDD-S optical
band light curves are systematically brighter than those of the HeD-S
model). Compared to the HeD-S model, the ELDD-S light curves reach peak
slightly earlier (and are a few tenths of a magnitude brighter at
peak). This follows from the different velocity distribution of the products of
He burning in these models.
In the ELDD-S model, the He-layer ashes have all been pushed out to relatively high velocity by the underlying core material (see Figure~{\ref{fig:compositions}}). Consequently, the outward column density from the surface 
layer of $^{56}$Ni and $^{52}$Fe is smaller in the ELDD-S model, leading to earlier $\gamma$-ray escape and downturn of the \textit{UVOIR} light curves. This 
effect, however, is relatively modest in scale: overall the optical band ELDD-S light curves are not very different from those of the HeD-S model. Similar conclusions can be drawn by comparing the HeD-L and ELDD-L optical band light curves in Figure~\ref{fig:lcsL}.

An important difference between ELDD and HeD light curves manifests in the near-infrared (e.g.\ the $J$-band for our S-model shown in Figure~\ref{fig:lcs}). 
In the near-infrared, a significant fraction of the emission in both the CSDD and ELDD models is provided by the intermediate-mass elements in the relatively cool and dense ejecta from the CO core, particularly around maximum light for the CSDD-S and ELDD-S simulations. {This emission from the core is powered by a combination of energy injected by the radioactive material in the core ejecta (significant for the CSDD-S model) and irradiation by the overlying He-shell ejecta -- around maximum light, re-radiation of energy originating from the He-detonation ash is the dominant source of NIR emission for the ELDD-S simulation.}
Since {only a very small mass of material was unbound from the CO core in our HeD models}, these processes are largely absent in the HeD models, making them fainter at these wavelengths\footnote{{Although the remnant CO WD will remain in the centre of the ejecta, it is expected to be too small to intercept a significant fraction of the radiation created in the rapidly expanding He detonation ash. Therefore, in contrast to the core ejecta in the CSDD and ELDD models, the WD is not expected to provide an effective target for re-radiating a significant fraction of the emission around maximum light.}}. Thus, near-infrared data could be particularly valuable when hunting for direct observable signatures of the core detonation.

\subsection{Colours and spectra}

Maximum light colours only weakly discriminate 
between our models. In all cases, $B-V$, $V-R$ and $V-I$ are positive around maximum light (see Figure~\ref{fig:colours} for our S-model results) and differ by at most a few tenths of a magnitude between the explosion scenarios. 

After maximum light, the $B - V$ colour rapidly becomes more positive because of the decline in $B$-band for all models. 
Evolution of the redder optical colours is complex but qualitatively similar to the colour evolution found in the double-detonation models studied by \citet{kromer10}. In particular, our CSDD-S model displays similar colour evolution to the lowest mass models in figure 3 of \citet{kromer10}. 
The ELDD-S and HeD-S models also show evolution to redder colours immediately after maximum light. This is both faster and more pronounced than in the CSDD-S model and both models show extremely red $V - I$ colours within two weeks of maximum light, a consequence of strong cooling emission by the Ca~{\sc ii} infrared triplet contributing to the $I$-band\footnote{Some caution must be applied to the interpretation of our prediction of very powerful Ca~{\sc ii} emission at ${\sim}30$~days after explosion in the ELDD and HeD models -- by these epochs the ejecta are sufficiently dilute that forbidden line emission may contribute significantly to the line cooling. Such emission is neglected in the current implementation of {\sc artis} \citep{kromer09} meaning that the calculations may overestimate the strength of Ca~{\sc ii} emission at late epochs.}.

\begin{figure*}
\epsfig{file=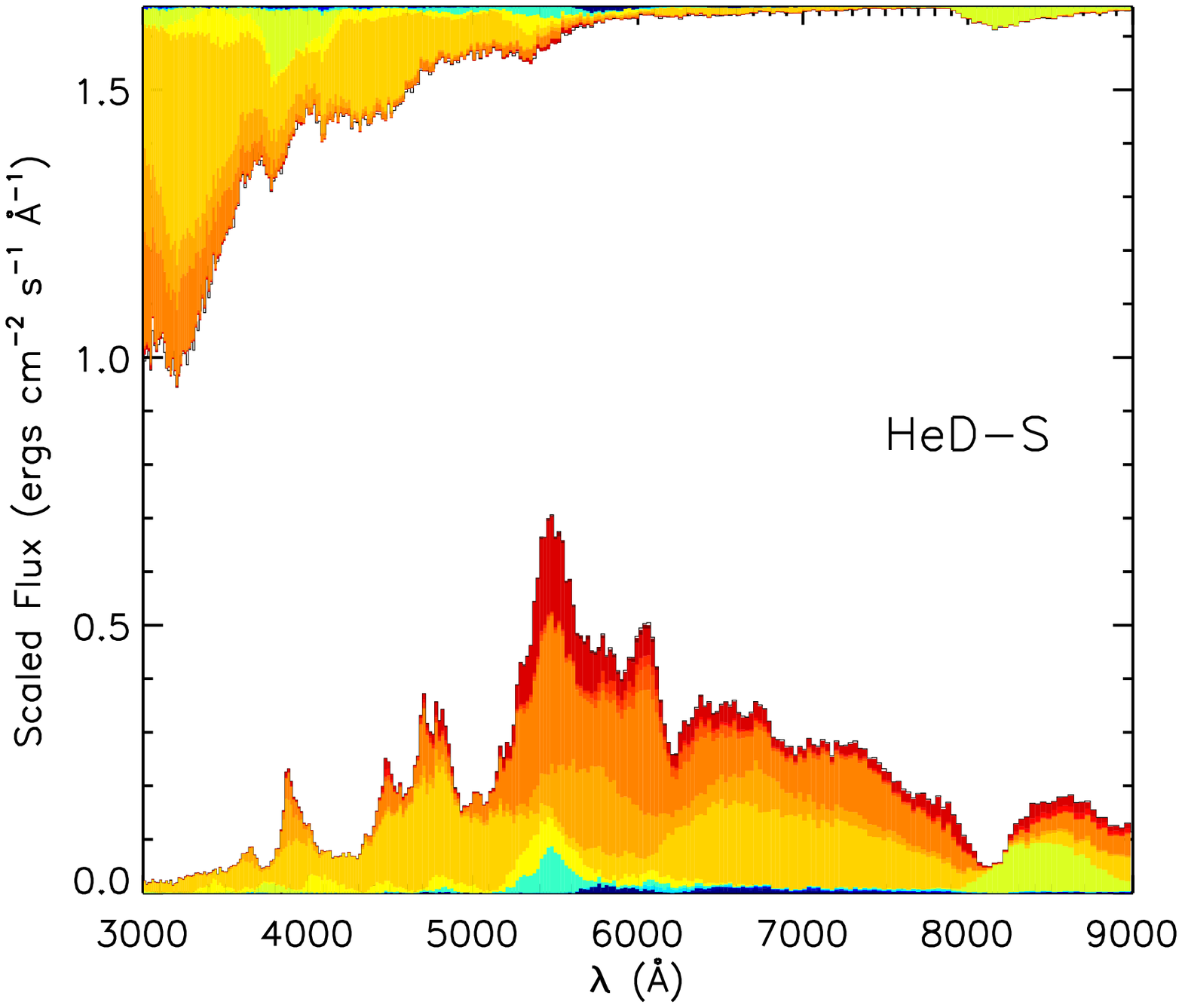,angle=90,width=8cm}
\epsfig{file=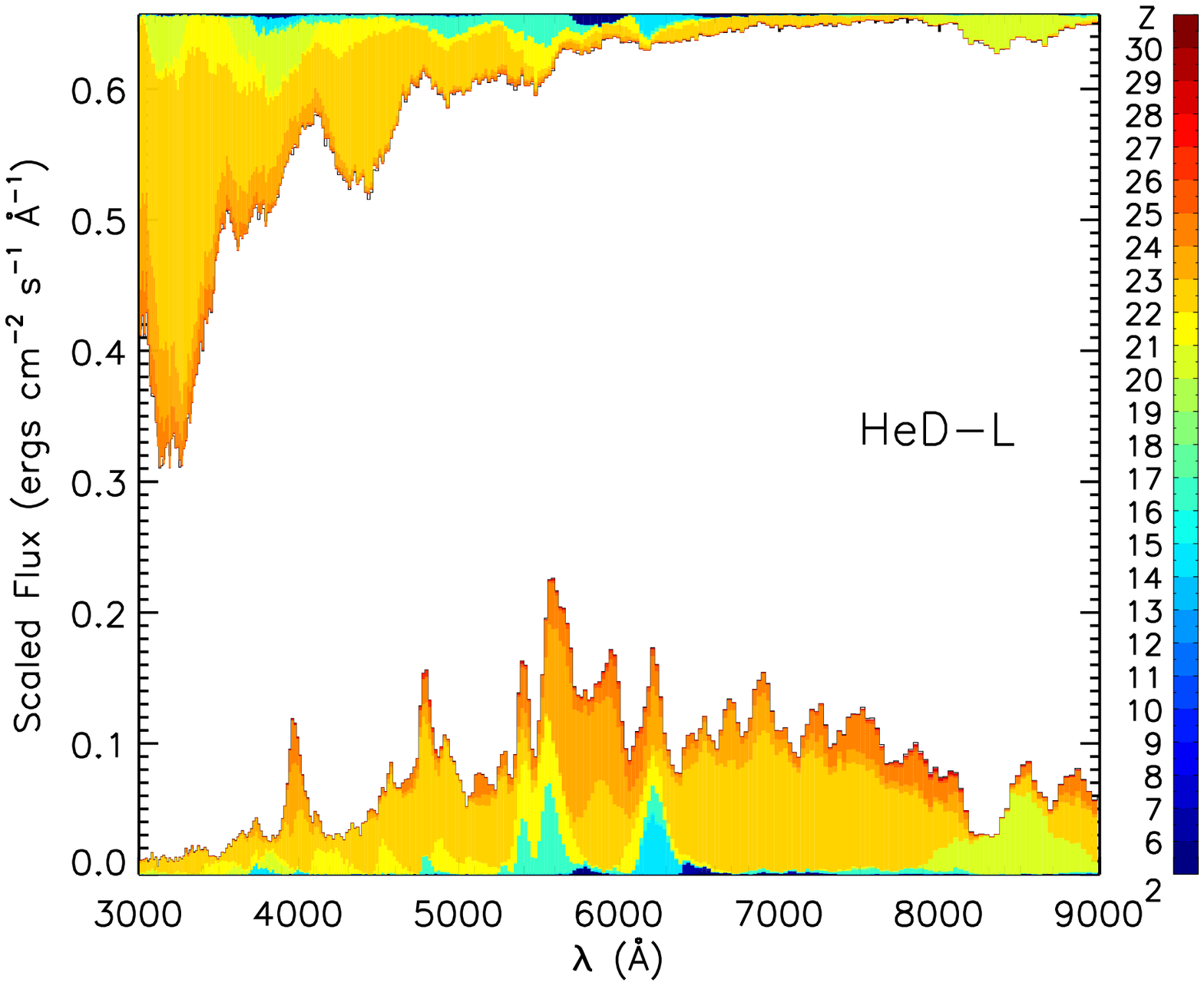,angle=90,width=8cm}
\caption{Optical spectra of model HeD-S (left) and HeD-L (right) at 8~days after explosion. The upper boundary of the coloured loci along the bottom of the plots are the synthetic spectra. The colour coding under the spectra identifies the elements with which escaping Monte Carlo quanta in each wavelength bin last interacted in our radiative transfer simulation. The coloured region along the top of the plots indicates which elements were last responsible for removing Monte Carlo packets from a particular wavelength bin (specifically, it shows the distribution of photon wavelengths that escaping packets had \emph{prior} to their last interaction). The colour bar on the right indicates the colour coding used for each atomic number ($Z$).}
\label{fig:spec_elem}
\end{figure*}

As in the \citet{kromer10} calculations, the red colours in our models are a consequence of effective line-blocking at blue wavelengths by iron group elements in the outer ejecta. In particular, the He detonation in all models yielded significant masses of Ti and Cr that, along with Ca, strongly influence the spectrum. This is illustrated in Figure~\ref{fig:spec_elem}, which shows the spectra of the HeD-S and HeD-L models at 8~days after explosion (the result is very similar for our other models). The colour coding in the figure indicates which elements were responsible for the last physical interactions of escaping Monte Carlo quanta in our radiative transfer simulations, making clear the dominance of elements with $Z=20$ to 24 (Ca to Cr) in shaping the emergent spectrum.

\begin{figure*}
\epsfig{file=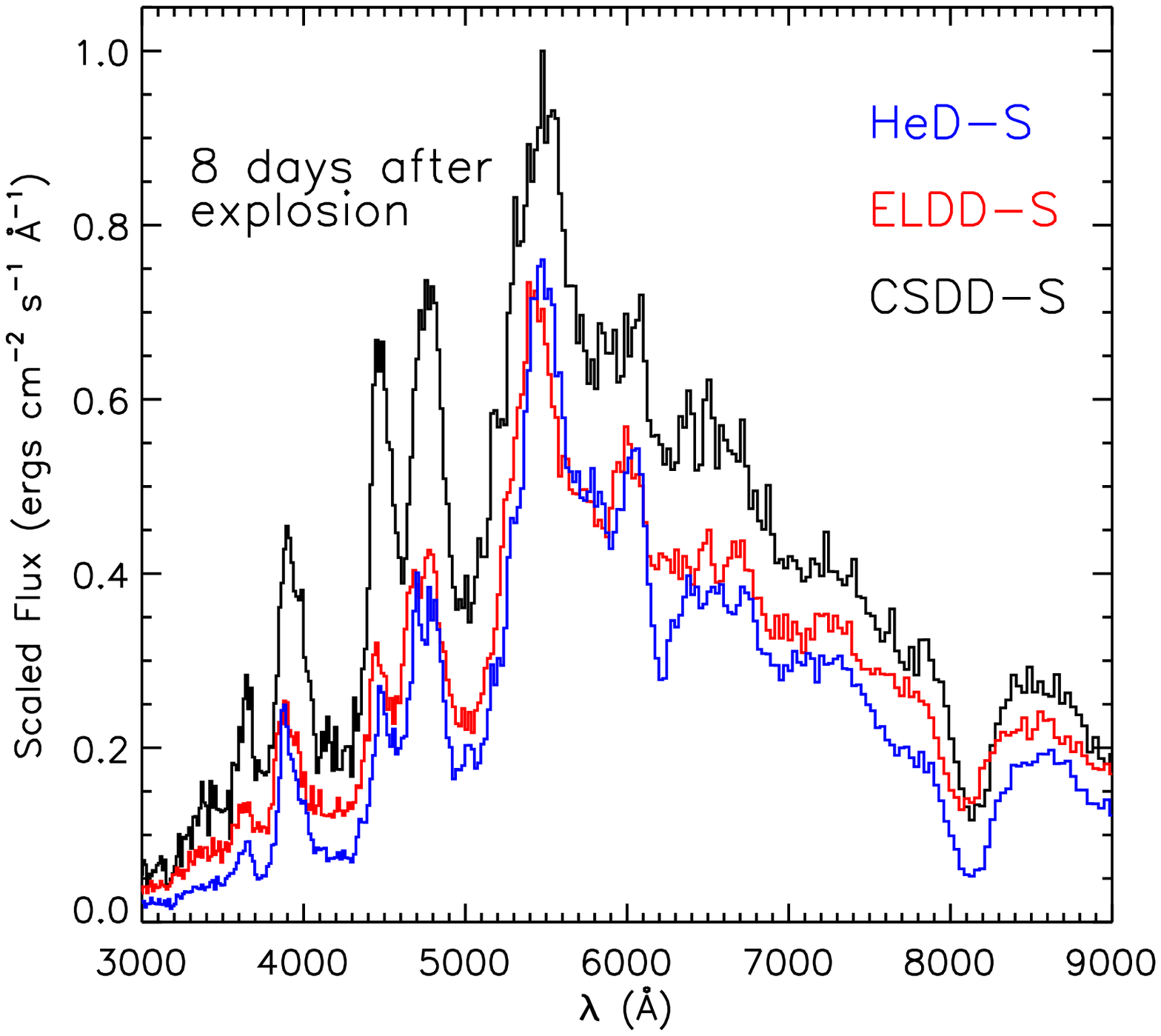,angle=90,width=8cm}
\epsfig{file=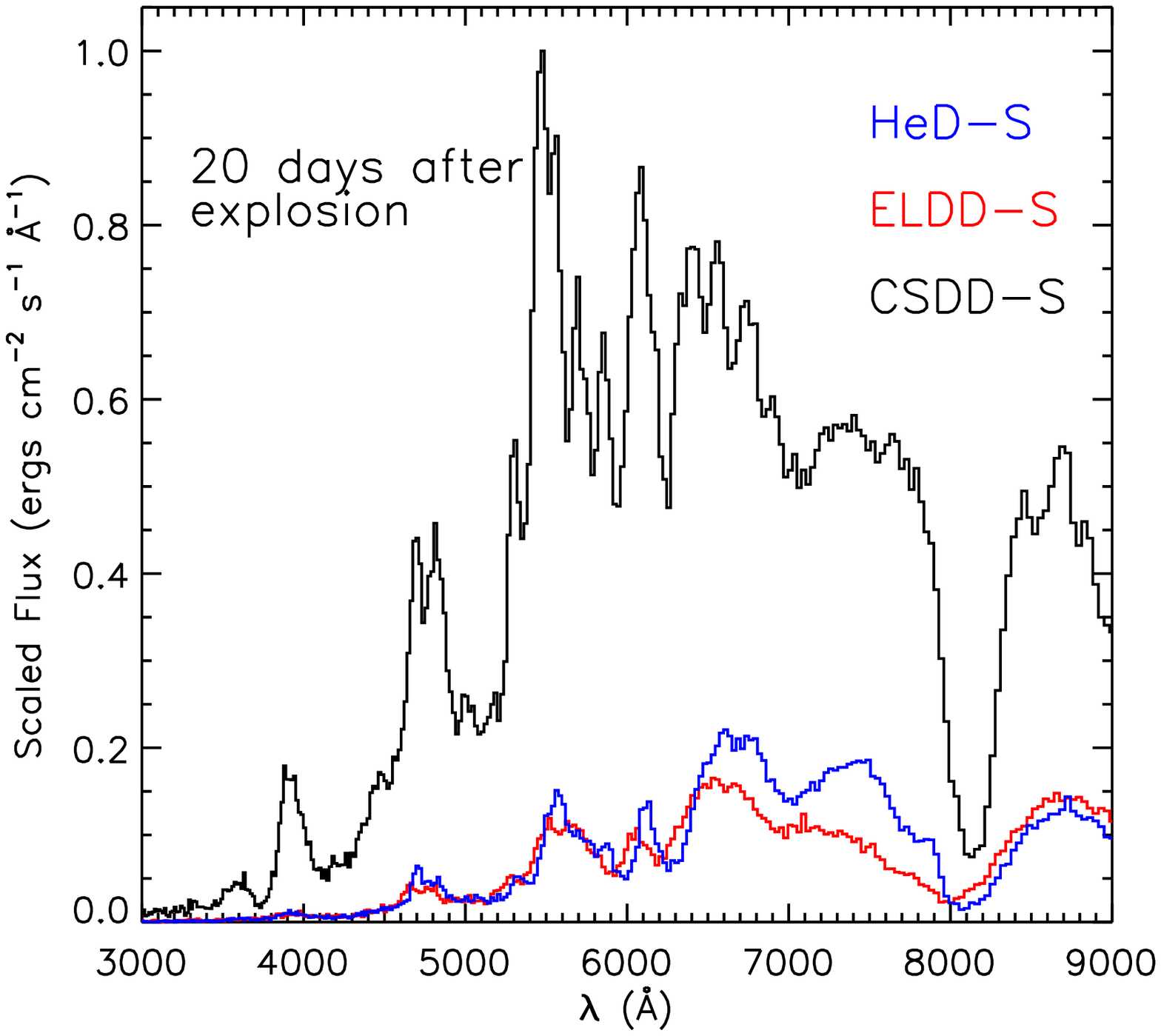,angle=90,width=8cm}\\
\epsfig{file=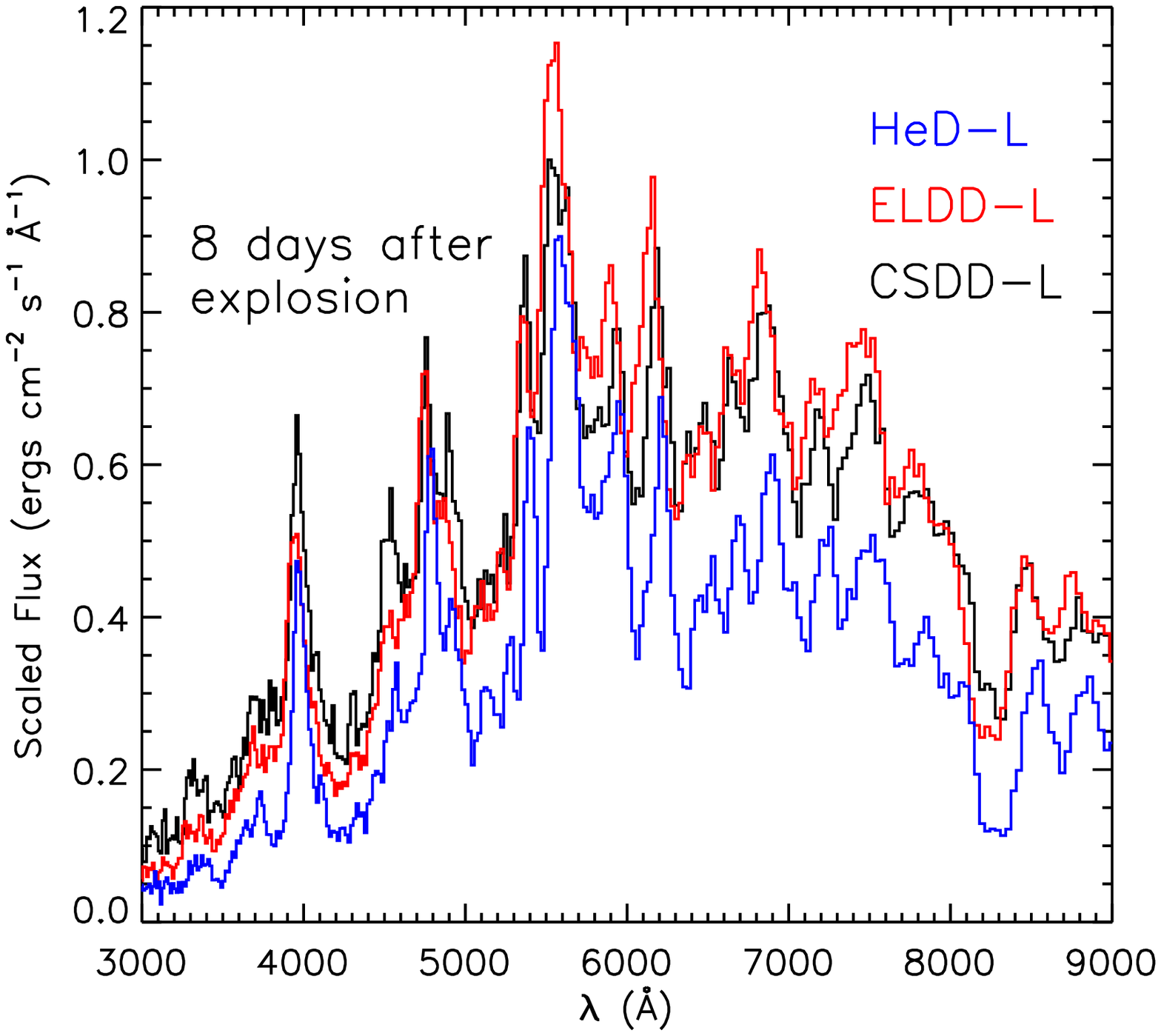,angle=90,width=8cm}
\epsfig{file=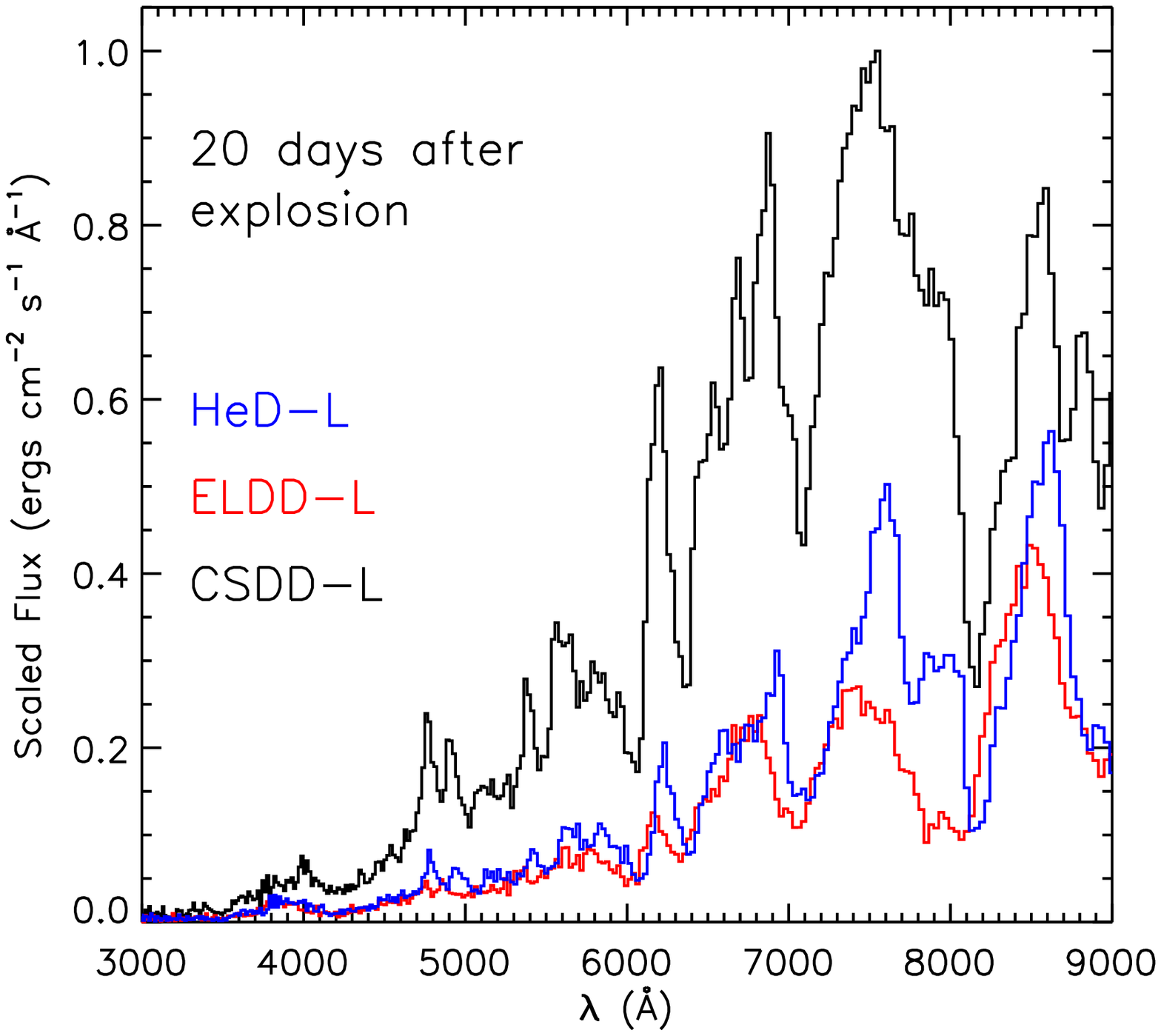,angle=90,width=8cm}
\caption{Optical spectra for our S- and L-models (top and bottom, respectively; CSDD in black, ELDD in red and HeD in blue) at 8 and 20~days after explosion (left and right panels, respectively). The flux scale is normalised to the peak of the CSDD model in all panels.}
\label{fig:spec_comp}
\end{figure*}

Figure~\ref{fig:spec_comp} compares the spectra for our explosion scenarios at
two epochs, 8 and 20~days after explosion. At the earlier epoch
(around peak brightness for the ELDD and HeD models), the spectra are
all quite similar with strong absorption in the Ca~{\sc ii} infrared
triplet and Ti~{\sc ii} lines (e.g.\ the characteristic trough around
4200~\AA). 
There are some subtle differences, however -- for example, the ELDD calculations (for both S- and L-models) typically show higher velocity line features than the HeD simulations, a consequence of the He-layer ejecta having been pushed to higher velocity by the CO detonation in the ELDD model. 
In general, our spectra are qualitatively similar
to the maximum light spectra for the low-mass models of
\citet{kromer10} but with a noticeable reduction in the role played by
silicon and sulphur.

By 20~days, the differences between the models are more apparent in the spectra. In particular, the CSDD-S model begins to show additional, relatively narrow line features (e.g.\ between 5600 and 5900~\AA) -- these are formed in the slowly expanding ejecta from the CO core detonation. In contrast, the ELDD-S and HeD-S spectra remain dominated by broad line features that form in the outer ejecta. In addition, the ELDD-S (and ELDD-L) spectra now very clearly show higher velocity features than the corresponding HeD models.

\subsection{Observer inclination}
\label{sect:viewing_angle}

\begin{figure*}
\epsfig{file=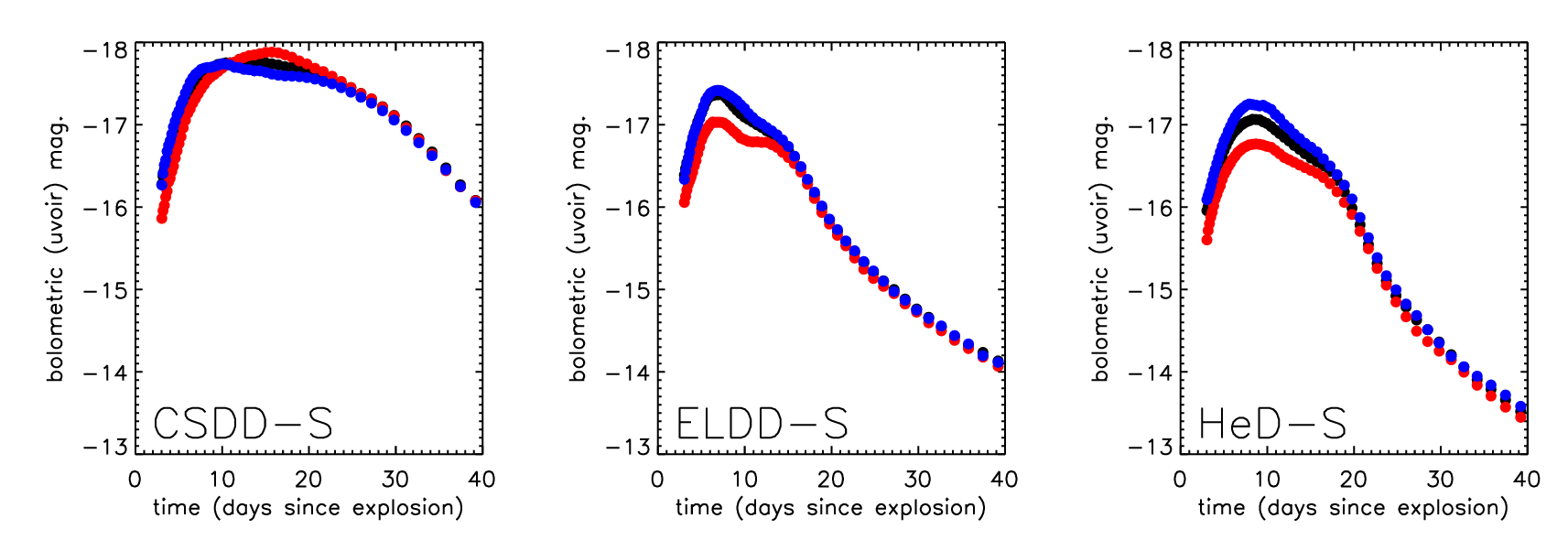,angle=90,width=15cm}\\
\epsfig{file=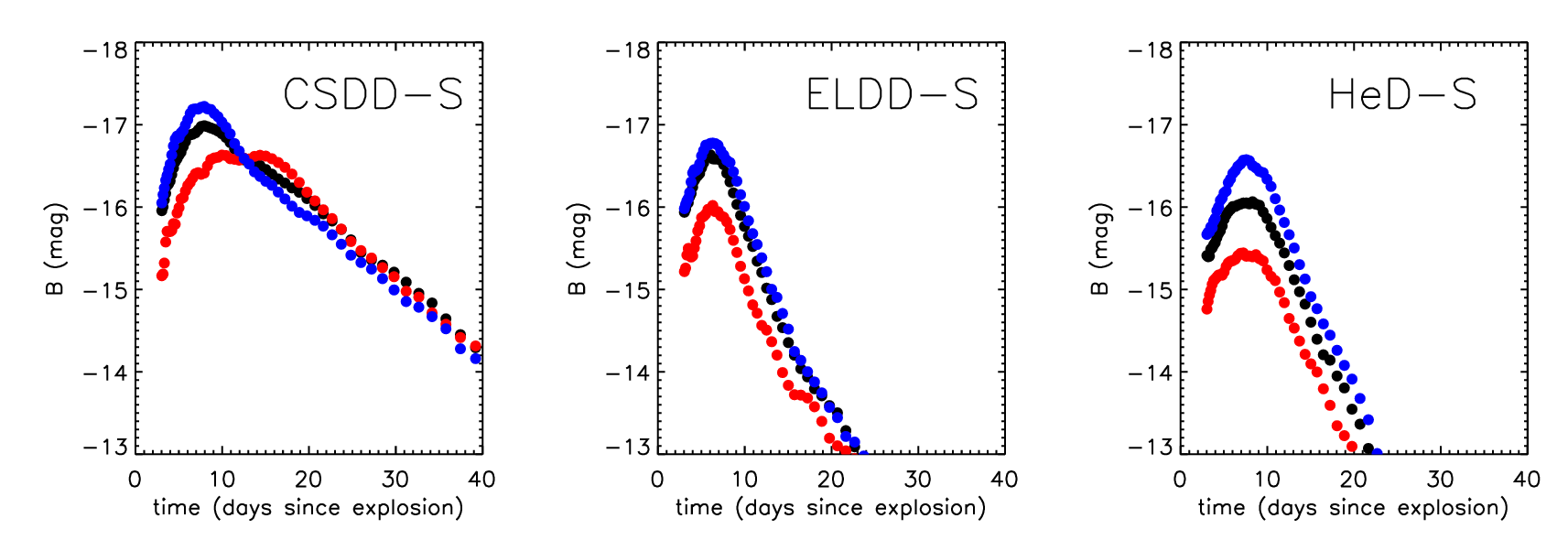,angle=90,width=15cm}\\
\epsfig{file=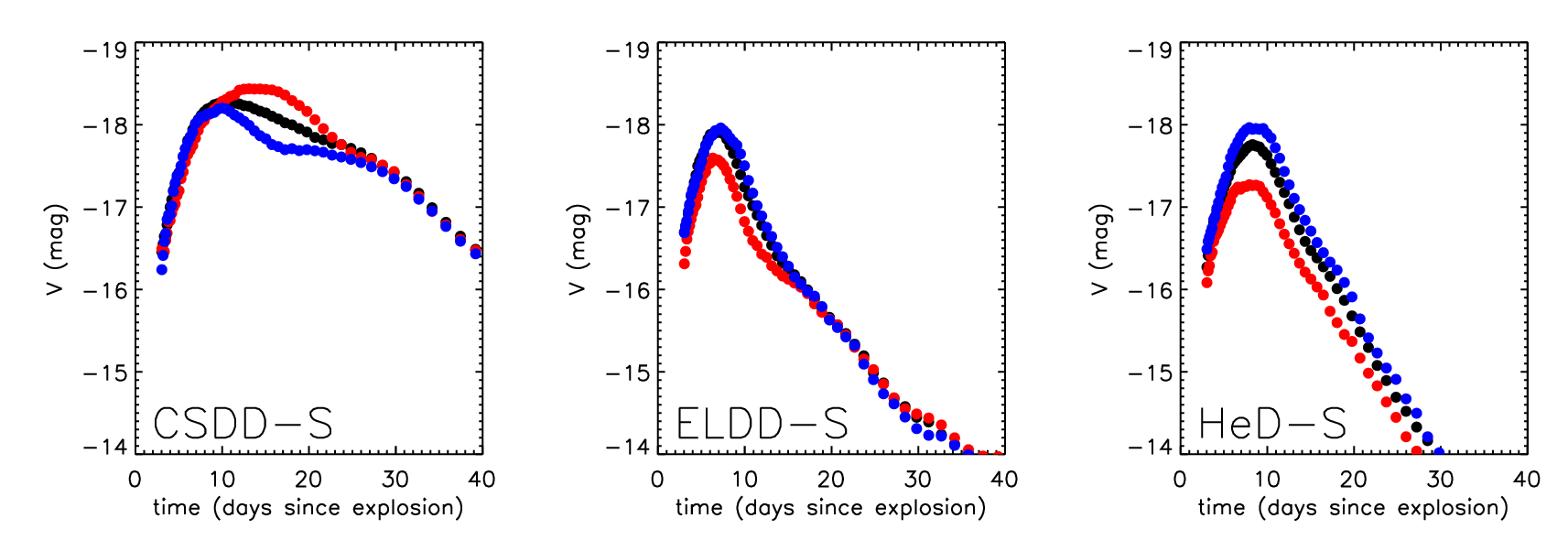,angle=90,width=15cm}\\
\epsfig{file=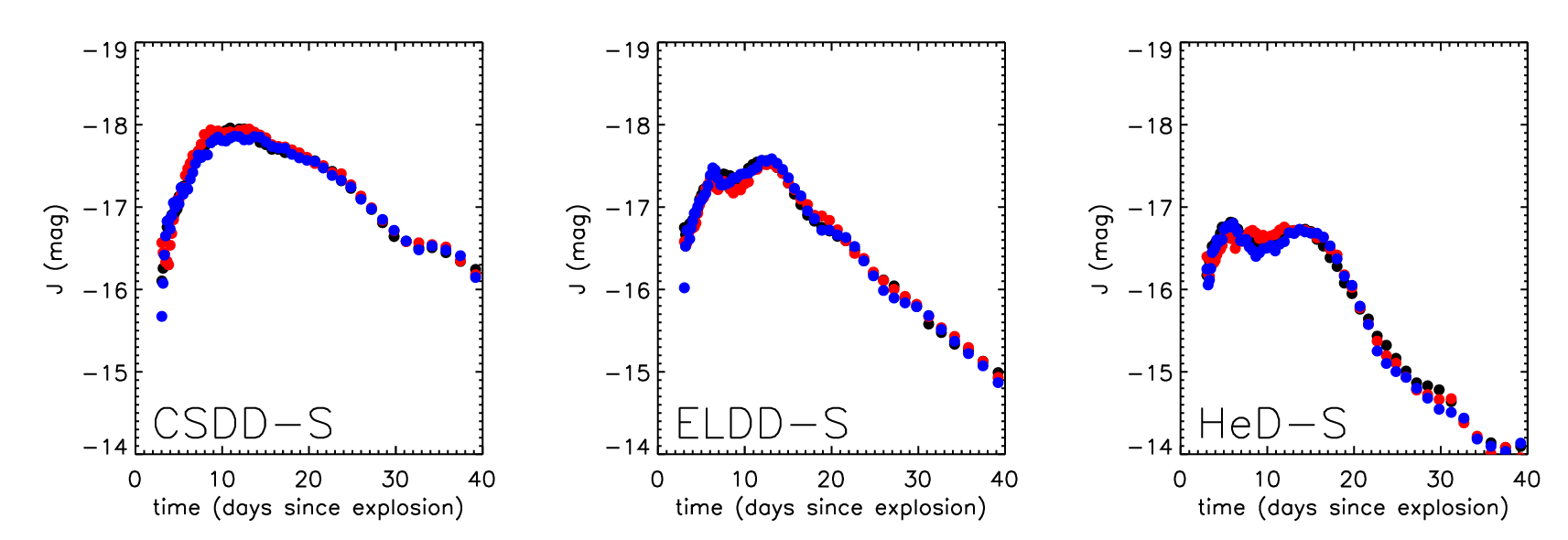,angle=90,width=15cm}
\caption{Light curves for selected observer inclinations to our CSDD-S, ELDD-S and HeD-S explosion models (left to right, respectively) in \textit{UVOIR}, $B$-, $V$- and $J$-bands. In each panel three light curves are shown for different observer orientations: viewed down the equator (black), from the $+z$-direction (blue) and the $-z$-direction (red). In all models the He ignition spot was on the $+z$-axis and the angle-averaged light curves (see Figure 3) are very similar to those for an equatorial line of sight.}
\label{fig:asym}
\end{figure*}

As discussed by \citet{fink10}, single-spot ignition of the He layer leads to a global asymmetry in the explosion that influences the ejecta from both the He layer and the core (see Figure~\ref{fig:compositions}). In particular, the He-layer detonation ash tends to sweep around the CO material (in the same sense as the laterally propagating He detonation), leading to a more geometrically extended layer of He-burning products around the pole opposite to the He ignition point.
This asymmetry affects the light curves and spectra, particularly at blue wavelengths \citep[see e.g.][]{kromer10}. 
Figure~\ref{fig:asym} illustrates this for our CSDD-S, ELDD-S and HeD-S models in the \textit{UVOIR}, $B$-, $V$- and $J$-band light curves. The \textit{UVOIR} curves are sensitive to observer inclination by several tenths of a magnitude while the bluer optical bands ($B$- and $V$-band) are affected more strongly (up to $\pm 0.5$~mag variation around the angle-averaged in $B$-band). As in the simulations discussed by \citet{kromer10}, the colours are bluest and the light curves decline most rapidly when viewed from the side on which the He detonation was ignited (i.e.\ from the $+z$-direction). 
At red wavelengths, observer inclination is less important and becomes negligible in the near-infrared (see lowest panels in Figure~\ref{fig:asym}), in accordance with the findings of \citet{kromer10}.

The influence of observer orientation is more complex in the CSDD-S model than the HeD-S and ELDD-S models. In that case, maximum light in both $B$- and $V$-band occurs significantly later (and is brighter in $V$) when viewed from the $-z$-direction. The broad light curve peaks in these models are sustained by radiation diffusing out from the $^{56}$Ni-rich inner parts of the ejecta (the CO ash). Since the CO detonation is off-centre (displaced down the $-z$-direction; see Figure~\ref{fig:compositions} and Table~\ref{tab:hotspot}), more of this radiation emerges in the $-z$-direction, leading to brighter extended optical emission when viewed from this direction \citep[cf.][]{sim07b}.

In the L-model simulations, the 
burning of the He-layer is significantly
more complete in the ash around the $-z$-axis than in other directions (the $^{56}$Ni mass fraction is small in most regions of the the He-layer ash but it becomes significant around the $-z$-axis because of the enhanced burning in the region where the He detonation converges\footnote{Due to reduction of surface area, the
detonation shock strengthens and the detonation becomes over-driven around the convergence point
\citep{livne1997a}. This affects the yields since the densities are just below the critical densities at which $^{56}$Ni is produced in a detonation.}).
Nevertheless, 
the effect of orientation on the L-model light curves is qualitatively similar to that found in the S-model and the scale of variation is comparable (up to $\pm 0.5$~mag in the blue bands and negligible in the near-infrared).

Overall, the influence of observer inclination is modest compared to the difference in light curve morphology between the CSDD and ELDD/HeD models. In particular, by ${\sim}20$~days after explosion, the light curves are fairly orientation-independent in all bands. Thus, observer orientation should not severely hinder observational discrimination between the CSDD and ELDD/HeD scenarios. However, it does affect the light curve shapes on a similar scale to the difference between the ELDD and HeD models and therefore will complicate attempts to distinguish these mechanisms.

\section{Discussion and conclusions}
\label{sect:discuss}

\subsection{Core detonation}
\label{sect:core_det}

In previous work, we have studied the double-detonation model applied to sub-Chandrasekhar mass CO white dwarfs with masses ${\sim}1$~M$_{\odot}$: this scenario predicts transients with a range of brightness compatible with SNe~Ia \citep[e.g.][]{hoeflich96a,nugent97,kromer10,woosley11} and appears to be a robust explosion mechanism \citep{livne95a,fink10}. 

Here we have investigated the plausibility of double-detonation models for
systems with relatively low CO core mass and high He-layer mass,
following the methodology of \citet{fink10}. We concluded that, if
detonation of the He layer occurs, the resulting compression and
heating of the CO core by inwardly propagating shocks can produce sufficiently 
high densities and temperatures that core detonation may occur, even for
core masses as low as 0.45~M$_{\odot}$. Together with previous results
\citep{livne95a,fink10} this suggests that detonation of an accreted
He layer (as in the p-Ia scenario of \citealt{bildsten07}) could be
accompanied by explosion of the underlying core for all CO core masses
that are commonly realized in nature.

Although our results suggest that core detonation is probable in all cases, they do not prove that it must always occur. For example, strong rotation might inhibit the shock convergence. Alternatively, the converging shocks might heat the material around the putative detonation point sufficiently that burning occurs prior to ignition of a detonation. Depending on the geometry, this might mean that the CO hot spot is completely enshrouded in nuclear ash, starving the detonation of fuel such that it is not able to propagate and incinerate the whole star. To investigate these possibilities in the future will require high-resolution, three-dimensional hydrodynamical/nucleosynthesis simulations. It will also be important to study the double-detonation mechanism for systems with ONe WD cores -- in this case it will be harder to ignite a core detonation, potentially opening an alternative parameter-space for He-layer explosions.

\subsection{Observable signatures of core detonation}
\label{sect:discuss_obs}

For large CO core masses ($\simgt 0.9$~M$_{\odot}$), the double-detonation model predicts light curves that are primarily powered by the large mass of $^{56}$Ni synthesised in the core detonation. Therefore these events will look very different from predictions of the p-Ia scenario for explosion of the corresponding He layer without the core detonation.
For lower mass CO cores, however, the $^{56}$Ni mass produced in the
core explosion is reduced such that some portion of the light curves
will be predominantly 
powered by the radioactive nuclei formed in the burning of the
He layer, as in the p-Ia model. To study this, we performed radiative
transfer simulations for three explosion scenarios: He detonation followed by core
detonation triggered following shock convergence (CSDD models), He detonation followed by prompt edge-lit core detonation (ELDD models) and He detonation with no core detonation (HeD models).
We investigated these explosion scenarios for two initial model systems ($M_{\mbox{\scriptsize CO}} = 0.58$ and $0.45$~M$_{\odot}$ together with He layers of $0.21$~M$_{\odot}$).

We found that the CSDD models can be easily distinguished from the ELDD and HeD models by their light curve morphologies: compression of the core prior to detonation means that the CSDD model yields relatively large masses of $^{56}$Ni from the core (for both our initial systems, the mass of $^{56}$Ni produced in the core is comparable to the total mass of radioactive nuclei produced in burning of the He layer). This central $^{56}$Ni causes the light curves to fade more slowly than in the other models. Consequently, this scenario can be easily distinguished observationally from an explosion of a He surface layer alone.

In contrast, our ELDD models show that if core detonation is ignited
without significant pre-compression, its influence on the observables
is much more subtle for the low-$M_{\mbox{\scriptsize CO}}$ systems we consider. 
The modest differences between the synthetic observables for our ELDD and HeD 
models can mostly be attributed to the differing structure of the ejecta layers produced in the He detonation. In particular, explosion of the core in the ELDD model clears out the He detonation products from the inner ejecta leading to light curves that evolve a little more rapidly and spectroscopic features that are broader and more blue-shifted. 
Thus rapidly evolving thermonuclear transients can be produced by edge-lit double-detonation of low-mass systems. Distinguishing them from pure He-layer detonations could best be done via the evolution of spectral line shapes and infrared photometry (see Section~\ref{sect:obs}).

It is noteworthy that the difference between our CSDD and ELDD models is stronger for our more massive initial system ($M_{\mbox{\scriptsize tot}} = 0.79$~M$_{\odot}$) than our extremely low-mass system ($M_{\mbox{\scriptsize tot}} = 0.66$~M$_{\odot}$). In both cases, the pre-explosion CO densities are too low to yield large $^{56}$Ni-masses in a prompt detonation. However, for our more massive system, the compression by converging shocks is able to raise the density sufficiently to produce $^{56}$Ni in a fairly substantial fraction of the core (our CSDD-S model) -- this leads to the differences in shape between our CSDD-S and ELDD-S light curves. For the low-mass system, the same effect occurs but is less dramatic because the initial CO densities are so low that even with shock compression only a small fraction of the core is burned to $^{56}$Ni.
For even more massive systems ($M_{\mbox{\scriptsize tot}} \simgt 0.9$~M$_{\odot}$),  a large fraction of the CO fuel will be at sufficiently high densities to produce $^{56}$Ni \emph{regardless} of shock compression \citep[see e.g.\ table 1 of ][]{sim10}. Double detonations of such systems will therefore always produce significant masses of $^{56}$Ni from the core, meaning that the differences between the ELDD and CSDD scenarios should be relatively small for massive systems.
Thus the best opportunity to observationally distinguish between the ELDD and CSDD mechanisms will be in systems where the pre-explosion densities in the core are close to but below the critical densities at which $^{56}$Ni is produced in a detonation.

\subsection{Relation to known transients}

The goal of this study has been to investigate whether secondary core detonation is likely for low mass CO cores and to predict the influence of CO core detonation on synthetic observables. We have not yet conducted an exploration of parameter space as required to attempt a quantitative comparison with observations. We can, however, comment qualitatively on the relation of our synthetic observables to the properties of known astrophysical transients. For reference, we tabulate important light curve shape parameters (rise times and decline timescales) for our simulations in Table~\ref{tab:lc_times}.

\citet{perets10} {proposed} that {the Type~Ib event} SN~2005E {could} be attributed to an explosion of an accreted He-layer on a WD\@.
Its peak brightness suggests that only a few thousandths of a solar mass of $^{56}$Ni, $^{52}$Fe and $^{48}$Cr were produced, if such a model is applicable. 
{A p-Ia like explosion was also discussed as one possibility to explain
SN~2008ha \citep{foley09}; this event was even fainter than SN~2005E and spectroscopically different, showing low expansion velocities and no evidence of He.} 
In both cases, however, the rise time was estimated to be $\simlt 10$~days and the light curve decline parameter was $\Delta M_{15}^{B} \sim 2$~mag. These rapid timescales clearly invite proper comparison with models for He-layer explosions.
\citet{waldman11} attempted to model SN~2005E in the context of the p-Ia scenario. They investigated a model with $M_{\mbox{\scriptsize CO}} = 0.45$~M$_{\odot}$ and  $M_{\mbox{\scriptsize He}} = 0.2$~M$_{\odot}$ (their CO.45HE.2 model) and found that such a model might be able to account for the peak brightness of SN~2005E. {However, they also found that, although significantly faster than normal SNe~Ia, the decline rate of SN~2005E was slower than could be easily explained with a p-Ia model.}

SN~2005E \citep{perets10} was significantly fainter ($M_\text{Bol.} > -15$~mag at peak) than our models.
Although the parameters of our L-system are very similar to the \citet{waldman11} CO.45HE.2 simulation, our explosions are significantly brighter. The large difference in brightness stems from the fact that {$^{44}\text{Ti}$} is the most abundant radioactive nucleus produced in the CO.45HE.2 simulation while our HeD-L model predicts {$^{48}\text{Cr}$} to be dominant. Although only a modest shift in the mean nucleosynthetic yields, this strongly affects the brightness since the half-life of {$^{44}\text{Ti}$} is much longer than that of {$^{48}\text{Cr}$}. Modification of our nucleosynthesis to produce less complete burning (i.e.\ a lower mass of {$^{48}\text{Cr}$}) could be achieved by reducing the density of the He layer, although this is likely unphysical \citep{shen09,woosley11}. 
{Alternatively, significantly polluting the He layer with a heavier element such as carbon will alter the nucleosynthesis \citep{kromer10,waldman11}. For a sufficiently large initial mass fraction, adding carbon reduces the typical atomic weight of the burning products and leads to more intermediate-mass elements rather than iron-group material. Reduced yields of radioactive iron-group elements would make our models fainter and alter the spectral features.}
We have not investigated such possibilities here. 
Nevertheless, our simulations have relevance to the study of SN~2005E and {similar faint and fast} transients, {potentially including e.g.\ SN~2008ha} -- in particular, they show that if a He layer detonation in a low-mass system is followed by a core detonation triggered by converging shocks, the core material can significantly retard the post-maximum decline of the light curves. For example, in our p-Ia-like HeD-L model, the $B$-band decline during the 15 day period after maximum light is $\Delta M_{15}^B = 3.4$~mag while for our CSDD-L model, $\Delta M_{15}^B = 2.5$~mag (the scale of this effect is even more dramatic in our S-model simulations; see Table~\ref{tab:lc_times}). Thus, in future studies, it will be important to consider whether CSDD explosions of low-mass systems might be able to account for faint thermonuclear transients whose light curves decline more slowly than can be explained by p-Ia models. 

It has also been suggested that p-Ia explosions may account for a class of very rapidly evolving thermonuclear explosions that includes SN~2002bj \citep{poznanski10}, SN~2010X \citep{kasliwal10}, SN~1939B and SN~1885A \citep{perets11}. These events are considerably brighter than SN~2005E: for example, \citet{kasliwal10} estimate $M_{r} \approx -17$~mag for SN~2010X while SN~2002bj is around 1.5~mag brighter. Moreover, the light curves of these events have a very rapid post-maximum decline (for example, $\Delta M_{15}^B \approx 3.2$~mag for SN~2002bj; \citealt{perets11}). The rapid post-maximum decline of this class of object is inconsistent with our CSDD simulations, particularly for our brighter S-model. However, it may be compatible with the rapid decline in our ELDD (or HeD) models 
(see Table~\ref{tab:lc_times}). Therefore focused modelling may be warranted to properly investigate whether ELDD models could be applicable to this class of transients.

\begin{table*}
\caption{Light curve rise times and decline rate parameters for the
  models. $t_\text{\textit{UVOIR}, max}$ and
$t_{B, \, \text{max}}$ are the times after explosion to
maximum light in \textit{UVOIR} and $B$-band, respectively. $\Delta
M_{15}^\text{\textit{UVOIR}}$ and $\Delta
M_{15}^B$ parametrize the light curve decline
rate in \textit{UVOIR} and $B$-band light (specifically, they give
the increase in magnitude during the 15 days after maximum light 
in the appropriate light curve).}
\label{tab:lc_times}
\begin{tabular}{ccccccc}\hline
Parameter & CSDD-S & ELDD-S & HeD-S & CSDD-L  & ELDD-L & HeD-L\\ \hline
$t_\text{\textit{UVOIR}, max}$ (days) & 12 & 7.1 & 8.7 & 11 & 7.6 & 9.9\\
$\Delta M_{15}^\text{\textit{UVOIR}}$ (mag)& 0.4 & 1.8 & 1.8 & 1.3 & 2.0 & 1.9\\
$t_{B, \, \text{max}}$ (days) & 8.1 & 6.0 & 7.7 & 5.4& 5.2 & 6.4\\
$\Delta M_{15}^B$ (mag) & 1.2 & 3.2 & 3.1 & 2.5& 4.0 & 3.4
\\ \hline
\end{tabular}
\end{table*}

\subsection{Future work}
\label{sect:future_work}

There remain many open questions to be addressed. Most important,
perhaps, is detailed study of burning of the He layer. The He
burning products play a critical role in determining the observational
properties of double-detonation explosion models but their yields
are sensitive to many issues including the exact densities and
composition of the He layer
\citep{shen09,kromer10,woosley11,waldman11}, the direction in which
the detonation propagates (radial or azimuthal;
\citealt{fink10,woosley11}) and the structure of the detonation
front. For example, \citet{kromer10} and \citet{waldman11} both showed
how introducing C to the He-layer prior to burning can alter the final
composition -- for double detonation of models with massive CO cores,
this dramatically affects the colours and can lead to much improved
agreement with observed SNe~Ia spectra. 

In addition, we have shown
that the observational consequences of core detonation in low-mass CO
cores are quite different if the core detonation is edge-lit rather
than triggered by shock compression. 
Although the triggering of the secondary detonation in our CSDD models is based on previous studies of the necessary conditions for CO detonation \citep{niemeyer97,roepke07}, we have not investigated the physical plausibility of edge-lit detonation. 
Further study of this is clearly warranted.
This is of primary interest for models with low
CO core masses (e.g.\ the S-model studied here rather than those of
\citealt{fink10} and \citealt{kromer10}) since the difference between the CSDD and ELDD mechanisms will be largest in this case (see Section~\ref{sect:obs}).
However, it does have some relevance to the study of more
massive cores since the position of an
off-centre ignition imprints distinctive signatures on the explosion ejecta \citep{chamulak11}.

It will also be important to quantify the relative frequency of bright (i.e. SNe~Ia-like luminosity) and fainter (i.e.\ p-Ia-like luminosity) explosions as a check on the plausibility of the double-detonation scenario contributing to both populations. 
The core/He-layer mass combinations used in this study are close to the low-mass extreme
for which potential progenitor systems 
can be realised in nature. For low core masses, the minimum mass of accreted He required for detonation becomes large \citep{bildsten07}.
As noted by \citet{shen10}, systems with massive He layers ($\simgt
0.1$~M$_{\odot}$) are not expected to be reached via the evolution of
systems in which a CO WD accretes from a He WD donor.
Our models, however, might be realised in systems where the donor is a
He-burning star \citep{iben91,shen09}. When investigating potential progenitors for SNe~Ia, \citet{ruiter11} found that the population of potential double-detonation systems with He-burning star donors is sub-dominant but can provide a significant event rate in young stellar populations. If extended to less massive CO cores, such population synthesis studies could estimate the relative occurrence of progenitors for the double-detonation scenario leading to both bright and fainter thermonuclear transients, a prediction that can be tested by the current and future generations of wide-field transient surveys.

\section*{Acknowledgments}

This work was supported by the NCI National Facility at the Australian National University, the Deutsche Forschungs\-gemeinschaft via
the Transregional Collaborative Research Center TRR 33 ``The Dark
Universe'', the Excellence Cluster EXC153 ``Origin and Structure
of the Universe'' and the Emmy Noether Program (RO 3676/1-1). Parts of the 
simulations were carried out at the John von Neumann Institute for 
Computing (NIC) in J\"{u}lich, Germany (project HMU14).

SAS thanks Ivo Seitenzahl, Ken Shen, Lars Bildsten and Brian Schmidt for illuminating conversations. We thank R\"{u}diger Pakmor for developing tools to map nucleosynthesis tracer particle yields for our explosion models.

{We thank the referee for useful comments that helped clarify the presentation of our work.}

\bibliographystyle{mn2e}
\bibliography{snoc}

\label{lastpage}

\end{document}